\documentclass[epj]{svjour}
\usepackage{graphicx,mathrsfs,amsmath,bm}
\usepackage{color}
\newcommand{\gtorder}{\mathrel{\raise.3ex\hbox{$>$}\mkern-14mu
            \lower0.6ex\hbox{$\sim$}}}
\newcommand{\ltorder}{\mathrel{\raise.3ex\hbox{$<$}\mkern-14mu
            \lower0.6ex\hbox{$\sim$}}}

\begin{document}
\title{New class of hybrid EoS and Bayesian M-R data analysis}
\author{D. Alvarez-Castillo\inst{1,}\thanks{\emph{On leave from:} 
Instituto de F\'{i}sica, Universidad Aut\'{o}noma de San Luis Potos\'{i}, M\'{e}xico}
\and 
A. Ayriyan\inst{2} 
\and
S. Benic\inst{3,}\thanks{\emph{Present address:} University of Tokyo, Tokyo, Japan}
\and 
D. Blaschke\inst{1,4,}\thanks{\emph{Permanent address:} 
Institute of Theoretical Physics, University of Wroclaw, Wroclaw, Poland} 
\and
H. Grigorian\inst{2,}\thanks{\emph{Permanent address:} 
Department of Physics, Yerevan State University, Yerevan, Armenia}
\and
S. Typel\inst{5}
}                     
\offprints{}          
\institute{
Bogoliubov Laboratory of Theoretical Physics, JINR Dubna, Dubna, Russia
\and
Laboratory of Information Technologies, JINR Dubna, Dubna, Russia
\and 
Department of Physics, University of Zagreb, Zagreb, Croatia
\and
National Research Nuclear University (MEPhI), Moscow, Russia
\and
GSI Helmholtzzentrum f\"ur Schwerionenforschung GmbH, Darmstadt, Germany
}
\date{Received: date / Revised version: date}
%
\abstract{
We explore systematically a new class of two-phase equations of state (EoS) for hybrid stars that is 
characterized by three main features : 
(1) stiffening of the nuclear EoS at supersaturation densities due to quark exchange effects (Pauli blocking) between hadrons, modelled by an excluded volume correction,
(2) stiffening of the quark matter EoS at high densities due to multiquark interactions and
(3) possibility for a strong first order phase transition with an early onset and large density jump.
The third feature results from a Maxwell construction for the possible transition from the nuclear to a quark matter phase and its properties depend on the two parameters used for  (1) and (2), respectively. 
Varying these two parameters one obtains a class of hybrid EoS that yields solutions of the Tolman-Oppenheimer-Volkoff (TOV) equations for sequences of hadronic and hybrid stars in the mass-radius diagram which cover the full range of patterns according to the Alford-Han-Prakash classification following which a hybrid star branch can be either absent, connected or disconnected with the hadronic one.
The latter case often includes a tiny connected branch.
The disconnected hybrid star branch, also called "third family", corresponds to high-mass twin stars characterized by the same gravitational mass but different radii. 
We perform a Bayesian analysis and demonstrate that the observation of such a pair of high-mass twin stars would have a sufficient discriminating power to favor hybrid EoS with a strong first order phase transition over alternative EoS.  
} 
\maketitle
\section{Introduction}
\label{sec:intro}
The study of the internal composition of neutron stars is an active field of research which relies on astrophysical observations that allow to refine theoretical models. 
In this respect, the recent observations of massive neutron stars
\cite{Demorest:2010bx,Antoniadis:2013pzd} have imposed important constraints on the stiffness of the equation of state (EoS), and therefore on the density ranges
covered by the density profiles of such high-mass compact star interiors. 
On the other hand radius measurements are still far from being precise, with most of the methods relying 
on either indirect measurements or model dependent assumptions like, e.g., for the neutron star atmospheres. 
While there is a wide range of claimed radii starting from, e.g., $\sim 9$ km \cite{Guillot:2013wu}
to $\sim 15$ km \cite{Catuneanu:2013pz} we will proceed here on the assumption that the actual radii
are large, as reported by~\cite{Bogdanov:2012md,Hambaryan:2014}. 
For a recent review of astrophysical constraints on dense matter see, e.g., Ref.~\cite{Miller:2013tca}.  
 
In this contribution, we present a Bayesian analysis (BA) study case with a class of hybrid EoS characterized by two parameters.
The first one stands for the baryonic exluded volume that determines the stiffness of hadronic matter at 
high densities. 
The second one is the coupling strength for an 8-quark vector current interaction which regulates the stiffness of the high-density quark matter phase.
It turns out that within the range of variation for the excluded volume parameter there is a qualitative change in the mass-radius relation for high-mass stars: beyond a certain value for the excluded volume
the first order phase transition proceeds with a sufficiently large jump in the energy density to cause an instability which, thanks to the stiffness of the quark matter phase at high densities goes over to a stable sequence of hybrid stars, the so-called "third family" of  compact stars.
In this situation, the conditions are fulfilled for the high-mass twin phenomenon, where stars on the high-mass end of the second family of purely hadronic neutron stars are degenerate in mass with their  twin stars on the lower mass part of the third family of hybrid stars bearing a quark matter core, see  
\cite{Benic:2014jia} and references therein. 
For a recent classification of stable hybrid star sequences under generic conditions for the EoS, see
Ref.~\cite{Alford:2013aca}.
We note that the high-mass twin phenomenon is quite substantially based on a stiffening of both, the 
hadronic and the quark matter EoS towards higher densities; it is not obtained within a systematic 
parameter scan of hadronic vs. NJL quark matter EoS \cite{Klahn:2013kga} which lacks additional stiffening effects as those introduced in Ref.~\cite{Benic:2014jia}.

This possibility of high-mass twin stars is of great importance not only due to the possibility of identification of a critical endpoint in the QCD phase diagram~\cite{Blaschke:2013ana,Alvarez-Castillo:2015xfa} but also because it provides a solution to several issues discussed in~\cite{Blaschke:2015uva}: 
the hyperon puzzle~\cite{Baldo:2003vx}, the masquerade problem~\cite{Alford:2004pf} and the reconfinement case~\cite{Lastowiecki:2011hh,Zdunik:2012dj}. 
Moreover, the transition between twin stars bears an energy reservoir \cite{Alvarez-Castillo:2015dqa}
that qualifies it as a possible engine for most energetic explosive astrophysical phenomena like gamma-ray bursts and fast radio bursts or play a role in contributing to the complex mechanism of core collapse supernova explosions. 
The present BA in the restricted two-dimensional parameter space of the new hybrid EoS
performed with the modern high mass and large radius priors will give an answer to the question whether
the high-mass twin star case is preferable over the connected hybrid star branch alternative.


\section{New class of quark-hadron EoS for hybrid stars}
\label{sec:hybrid}
In this study we consider hybrid neutron stars that are composed of hadronic matter and might undergo a phase transition to quark matter in their cores if parameter values of the models physically allow for it. 
In this way both \textit{pure hadronic} and \textit{hybrid} star configurations with physical properties determined by 
fixed parameter values shall be faced against observational data for model descrimination. 
The EoS description is presented in the following subsections. 

\subsection{Hadronic EoS with excluded volume corrections}
The description of hadronic matter in terms of pointlike hadrons with appropriately chosen interactions has to be limited to the low density region where effects of the finite size of hadrons (due to their compositeness) can be neglected.
In the high density region, however, the quark substructure of nucleons requires antisymmetrization of the many-nucleon wave function w.r.t. quark exchange among them, leading to the Pauli blocking effect in the EoS. 
This effect is expected to happen as density increases beyond saturation density and can be viewed as a precursor of quark delocalization as the equivalent of the quark deconfinement transition~\cite{Ropke:1986qs} 
Pauli blocking will be intensified by the simultaneous partial chiral restoration~\cite{Blaschke:2016}. 
The resulting consequences for the EoS can be mimicked by adopting excluded volume corrections to any purely hadronic EoS.

The excluded volume correction is applied at suprasaturation densities and has the effect of stiffening the EoS without modifying any of the experimentally
well constrained properties below and around saturation $n_{\textmd{sat}}=0.16$ fm$^{-3}$, the density in the interior of atomic nuclei. We introduce the  
the available volume fraction $\Phi_N$ for the motion of nucleons at a given density $n$ as~\cite{Typel:2016srf}
 \begin{equation}
    \Phi_N=\left\{
                \begin{array}{lll}
                  1~,& \textmd{if} &n \leq n_{\textmd{sat}}\\
                  \exp[-{v\vert v \vert}(n-n_{\textmd{sat}})^{2}/2]~, & \textmd{if} &n > n_{\textmd{sat}}~,
                \end{array}
              \right.
              \label{vex}
  \end{equation}
with $v=16\pi r_N^3/3$ 
as the van-der-Waals excluded volume corresponding to a nucleon hard-core radius $r_N$.
The mathematical form of (\ref{vex}) with positive (negative) values for $v$ leads to a stiffening (softening)
of the original EoS.
We shall consider only positive values of excluded volume in this work and introduce for them the dimensionless parameter $p=10\times v[{\rm fm}^3]$, taking
values between $p=0$ and $p=80$.

For the hadronic part of the neutron star EoS we will consider here the density dependent relativistic meanfield  EoS named "DD2" \cite{Typel:2009sy} (which is rather stiff) and the slightly softer EoS "DD2F" which fulfills the flow constraint from heavy-ion collision experiments 
\cite{Danielewicz:2002pu}. 
This flow constraint version of the DD2 EoS is obtained by multiplying the density dependent meson-nucleon couplings 
$\Gamma_i(n)$ of the DD2 model \cite{Typel:2009sy}  for $n>n_{\rm sat}$ with the functions $g_i(n)$, where $i=\omega, \sigma$.  
These functions are defined as
\begin{equation}
g_\omega(n) = \frac{1+\alpha_- x^s}{1+\alpha_+ x^s} = \frac{1}{g_\sigma(n)}~,
\end{equation}
where $x=n/n_{\rm sat} - 1$ and $\alpha_\pm = k(1\pm r)$, with the parameter values being adjusted to
$k=0.04$, $r=0.07$ and $s=2.25$. 

For both these EoS we shall adopt a variation of the nuclear symmetry energy as introduced 
in~\cite{Typel:2014tqa} in the context of the description of neutron skin thickness of heavy nuclei. 
The symmetry energy $E_s(n)$ is defined as the difference in the energy per nucleon between pure neutron matter and symmetric matter in a uniform, infinite system. 
In RMF models the isovector $\rho$ meson usually represents the only contribution to the isospin 
dependence of the interaction. Following~\cite{Typel:2014tqa} we use here three parametrizations 
for the density-dependent $\rho$ meson coupling
\begin{equation}
\Gamma_\rho(n) = \Gamma_\rho(n_{\rm sat}) \exp (-\alpha_\rho x)~,
\end{equation}
corresponding to a "soft", "medium" and "stiff" symmetry energy, see Table~\ref{tab:rho}.

\begin{table}[!h]
\label{tab:rho}
\begin{tabular}{lllll}
\hline
\hline
$E_s(n)$ & \multicolumn{2} {l} {parametrization} & $\Gamma_\rho(n_{\rm sat})$ & $a_\rho$ \\
\hline
stiff & DD2$+$ & DD2F$+$ & 3.806504 & 0.342181\\
medium & DD2 & DD2F & 3.626940 & 0.518903\\
soft & DD2$-$ & DD2F$-$ & 3.398486 & 0.742082\\ 
\hline
\hline
\end{tabular}
\caption{Parametrization of the symmetry energy variation by different density dependences of the 
$\rho$ meson coupling, adapted from Ref.~\cite{Typel:2014tqa}.}
\end{table}

It turns out that $E_s(n)$ has a direct impact on the neutron star radius, in particular for the low-mass stars like star (B) of the double pulsar PSR J0737-3039(B) \cite{Kramer:2006nb}, thus affecting the compactness and binding energy, important observables for our analysis in this work. 
We have hereby defined a family of six EoS which we label according to their symmetry
energy functionals as DD2, DD2-, DD2+, DD2F, DD2F-, and DD2F+, respectively.
For each of these EoS we will vary the excluded volume as the first free EoS parameter of this study.

\begin{figure*}[!bht]
\begin{center}$
\begin{array}{ccc}
\includegraphics[width=0.25\textwidth]{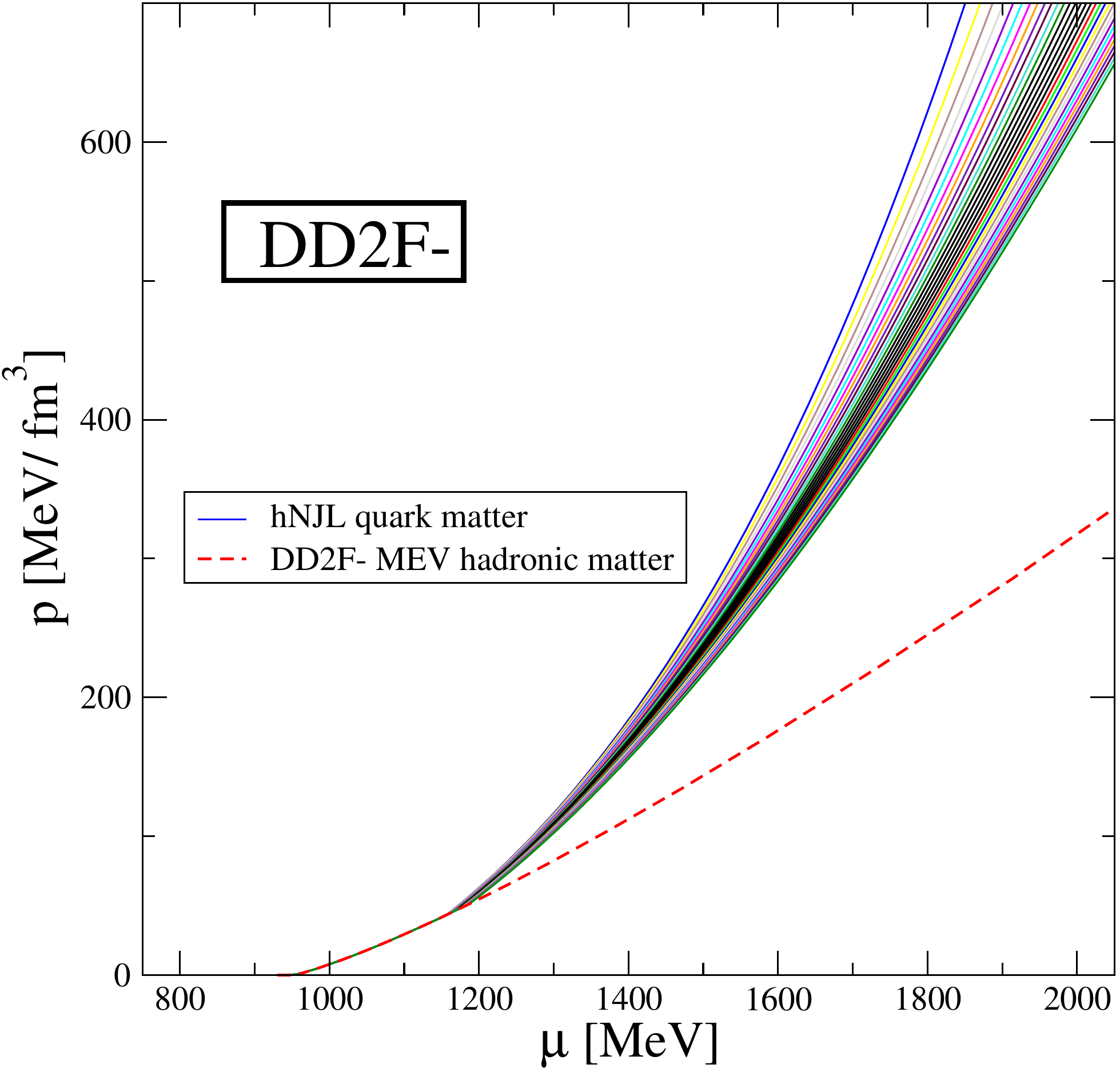} & \includegraphics[width=0.25\textwidth]{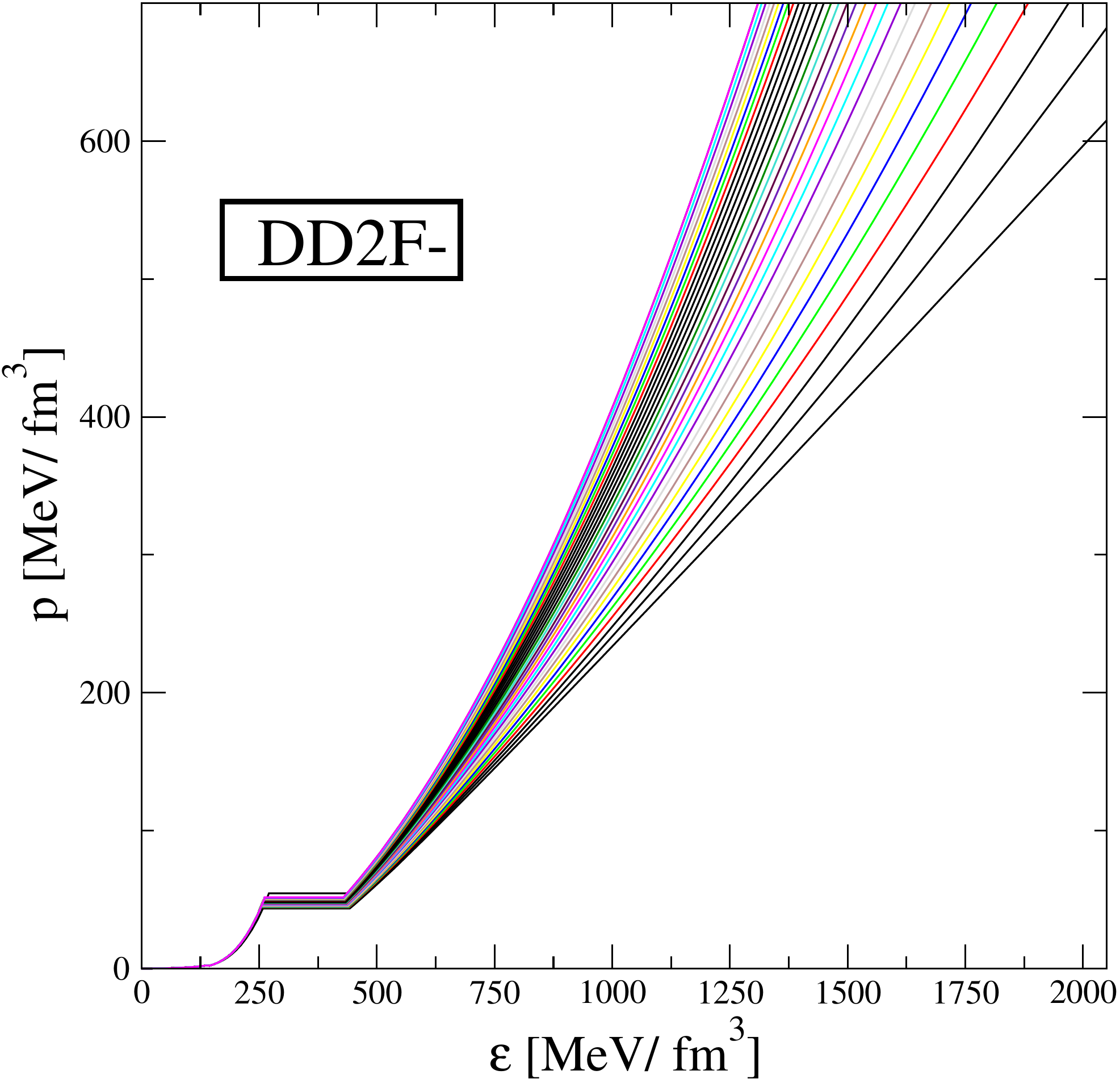} & \includegraphics[width=0.25\textwidth]{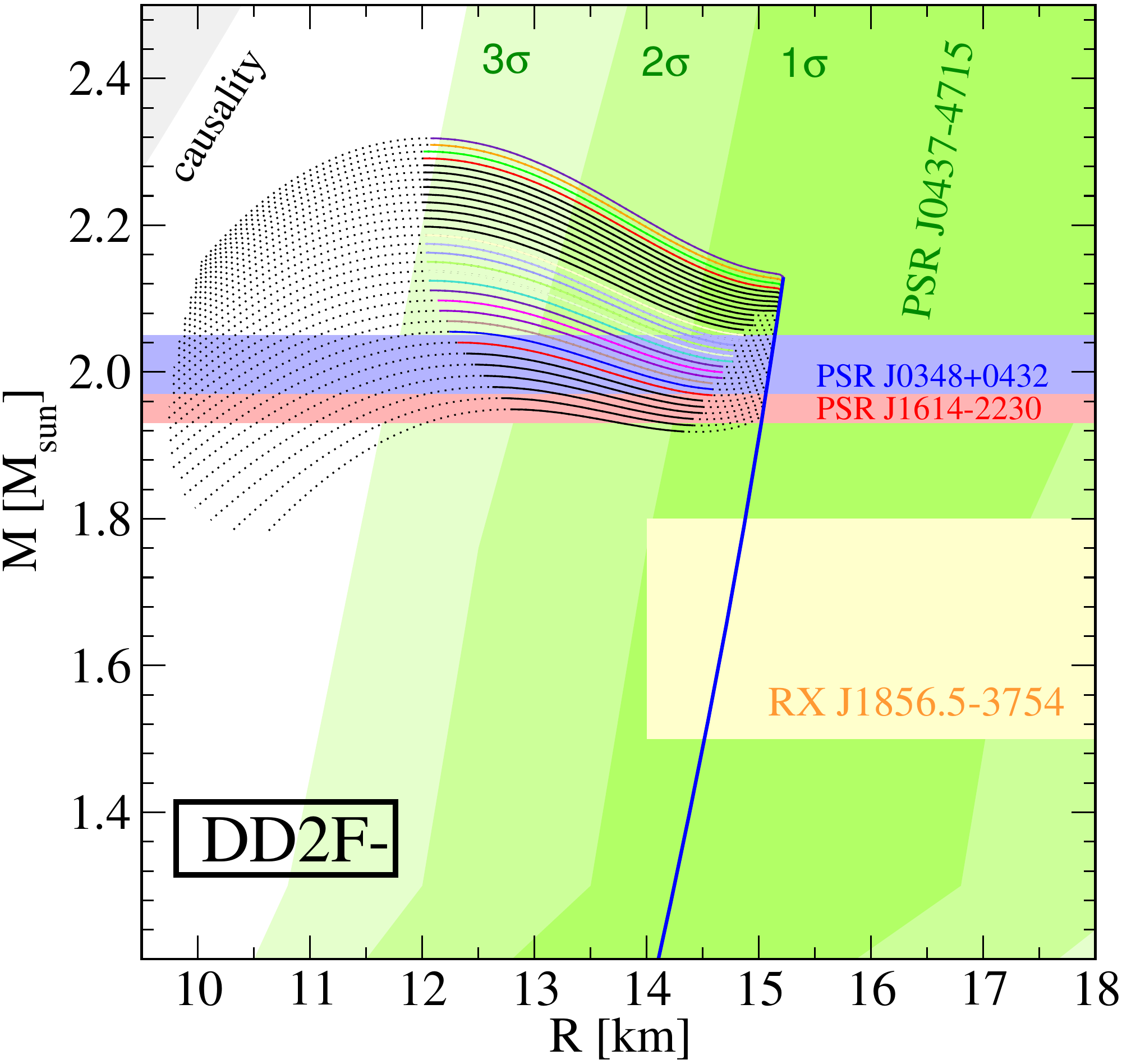}\\
\includegraphics[width=0.25\textwidth]{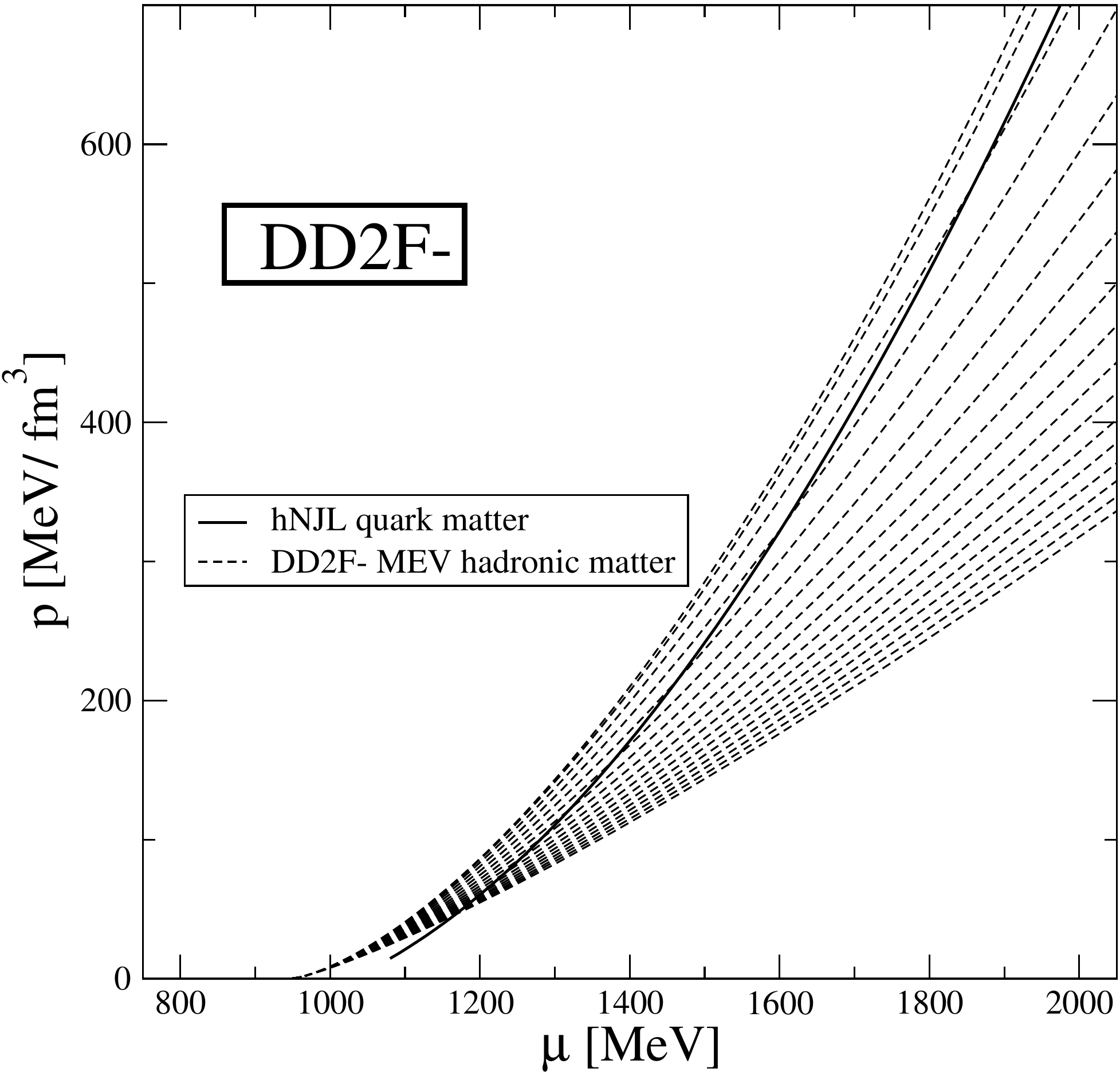} & \includegraphics[width=0.25\textwidth]{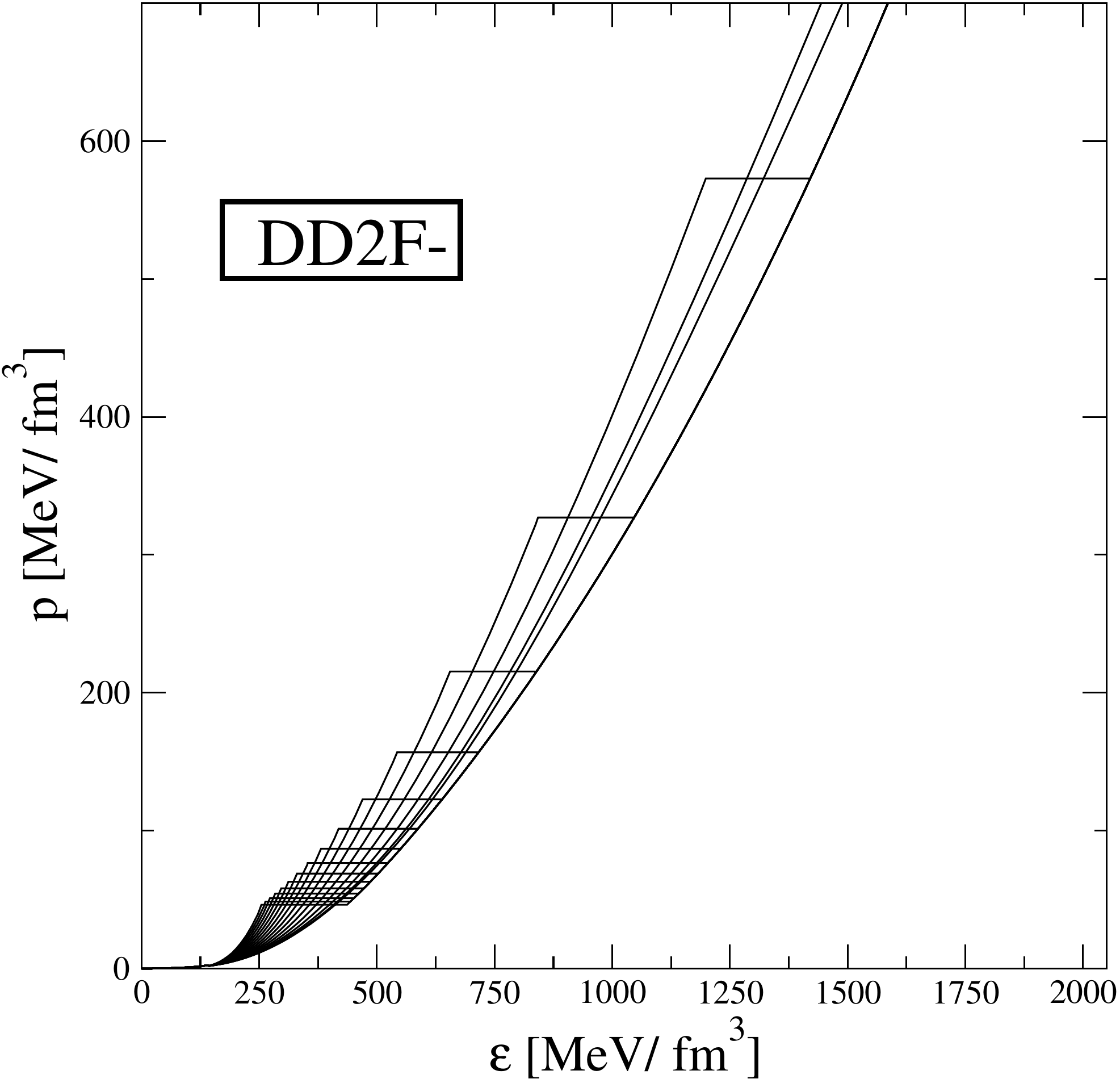} & \includegraphics[width=0.25\textwidth]{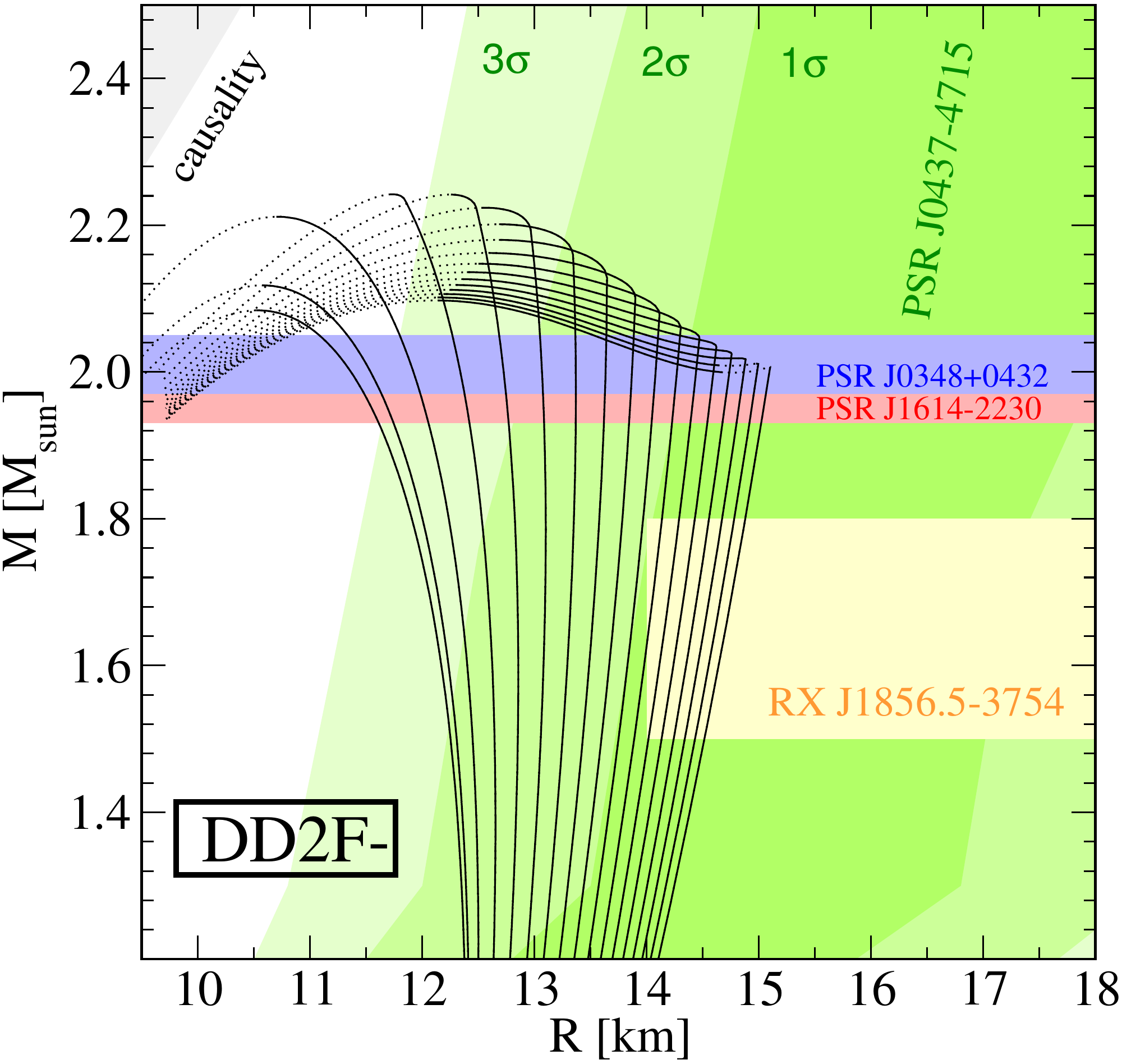}
\end{array}$
\end{center} 
\caption{Variations of the hybrid EoS for the DD2F$^-$ model. \textit{Upper row}. The hadronic EoS is kept fixed while the quark EoS is allowed to vary for the parameters $\eta_4=0,1,2..30$.
\textit{Lower row}. The quark EoS is fixed whereas the hadronic EoS takes the values $p=0,5,10,...80$. For all these models the EoS is shown on the left and central plots while the resulting mass radius diagrams are shown on the right side.}
\label{Case_A_n_B}
\end{figure*}

\subsection{Quark matter EoS with multi-quark interactions}

The EoS in the high density phase is obtained from a NJL model with multiquark interactions \cite{Benic:2014jia,Benic:2014iaa}.
The Lagrangian for two quark flavors, $q=(u,d)$ is defined as
\begin{equation}
\mathcal{L}=\bar{q}(i\partial_\mu\gamma^\mu-m)q+\mu_u\bar{u}\gamma^0 u+\mu_d\bar{d}\gamma^0 d +
\mathcal{L}_4+\mathcal{L}_8~,
\label{eq:njllag}
\end{equation}
where $\mu_f$ are the chemical potentials of flavor $q_f$ and $m$ is 
the current mass. 
The interaction terms are
\begin{eqnarray}
\label{eq:l4}
\mathcal{L}_4 &=& \frac{g_{20}}{\Lambda^2}
[(\bar{q}q)^2 + (\bar{q}\bm{\tau} q)^2]-\frac{g_{02}}{\Lambda^2}[(\bar{q}\gamma_\mu q)^2 + (\bar{q}\gamma_\mu\bm{\tau} q)^2]\,,\\
\mathcal{L}_8 &=& \frac{g_{40}}{\Lambda^8}[(\bar{q}q)^2 + 
(\bar{q}\bm{\tau} q)^2]^2-\frac{g_{04}}{\Lambda^8}[(\bar{q}\gamma_\mu q)^2+(\bar{q}\gamma_\mu\bm{\tau} q)^2]^2 \nonumber \\ 
&&-\frac{g_{22}}{\Lambda^8}[(\bar{q}\gamma_\mu q)^2+(\bar{q}\gamma_\mu\bm{\tau} q)^2][(\bar{q}q)^2 + (\bar{q}\bm{\tau} q)^2]\,.
\label{eq:l8}
\end{eqnarray}
While the NJL model has additional interactions to complete its chiral symmetry, in Eqs.~(\ref{eq:l4}) and (\ref{eq:l8}) only the components that condense at finite density are shown.
Within the mean-field approximation the thermodynamic potential is found
\begin{eqnarray}
\Omega &=& U+ \sum_{f=u,d}\Omega_f(M_f,T,\tilde{\mu}_f)-\Omega_0~,
\label{eq:thnjl}
\end{eqnarray}
where
\begin{eqnarray}
U &=& 2\frac{g_{20}}{\Lambda^2}(\phi_u^2+\phi_d^2) + 
12 \frac{g_{40}}{\Lambda^8}(\phi_u^2+\phi_d^2)^2 \nonumber \\
&&-2\frac{\eta_2 g_{20}}{\Lambda^2}(\omega_u^2 + \omega_d^2)-12 \frac{\eta_4 g_{40}}{\Lambda^8}(\omega_u^2+\omega_d^2)^2~,\\
\Omega_f &=&-2 N_c\int \frac{d^3 p}{(2\pi)^3}
\Big\{E_f + T\log[1+e^{-\beta(E_f-\tilde{\mu}_f)}] \nonumber \\
&&+ T\log[1+e^{-\beta(E_f+\tilde{\mu}_f)}]\Big\}~,
\end{eqnarray}
where we have set the mixing term $g_{22}=0$ \cite{Benic:2014iaa}.
The parameter $\eta_4$ is the dimensionless scaled coupling strength for the 8-quark interaction in the vector meson channel which determines the stiffness of the quark matter EoS at high densities.
It is the second parameter that will be subject to free variation in this study.

The quark pressure $P_q$ is obtained by solving the gap equations
\begin{equation}
\frac{\partial\Omega}{\partial \phi_f} = 0 ~, \quad 
\frac{\partial\Omega}{\partial \omega_f} = 0 ~.
\end{equation}
for the scalar ($\phi_f \equiv \langle \bar{q}_f q_f\rangle$) and vector 
($\omega_f\equiv \langle q^\dag_f q_f\rangle$) mean-fields
and evaluating (\ref{eq:thnjl}) as $P_q=-\Omega$.
$\beta$-equilibrium is maintained through the processes
$d\to u+e^- +\bar{\nu}_e$ and $u+e^- \to d + \nu_e$, so that
\begin{equation}
\label{beta-eq}
\mu_u = \mu_d + \mu_e , 
\end{equation}
where $\mu_e$ is the electron chemical potential.
We take into account charge neutrality by
\begin{equation}
\label{charge}
\frac{2}{3}n_u - \frac{1}{3}n_d - n_e = 0~,
\end{equation}
where $n_f = -\partial \Omega/\partial \mu_f$.

\begin{figure*}[!thb]
\begin{center}
\begin{tabular}{l|c|c|c}
\hline
Symmetry energy $\rightarrow$&&&\\
&soft & medium & stiff\\
EoS $\downarrow$&&&\\
\hline
&&&\\ 
DD2F (semi-soft)&
\includegraphics[width=0.25\textwidth]{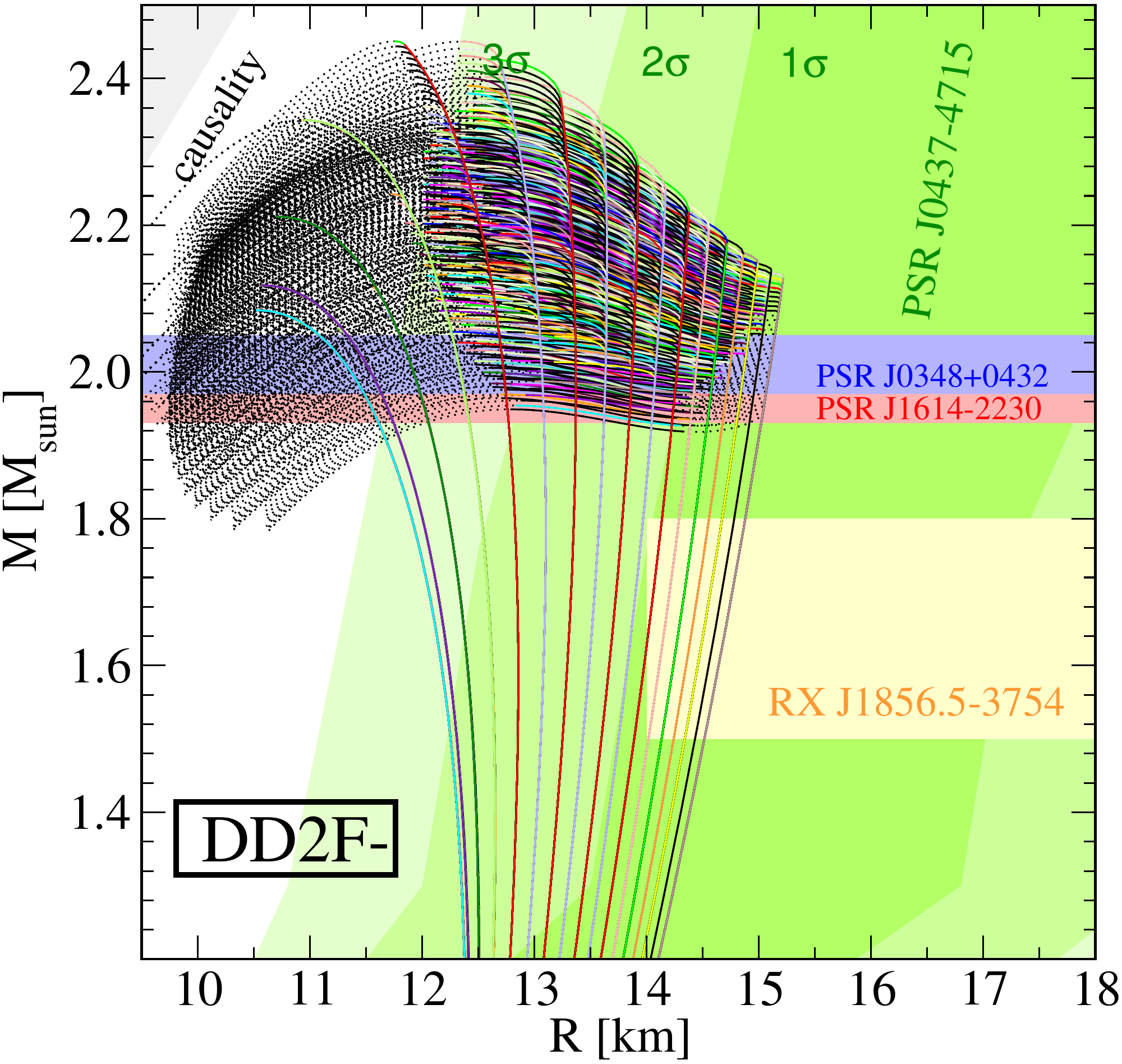} & \includegraphics[width=0.25\textwidth]{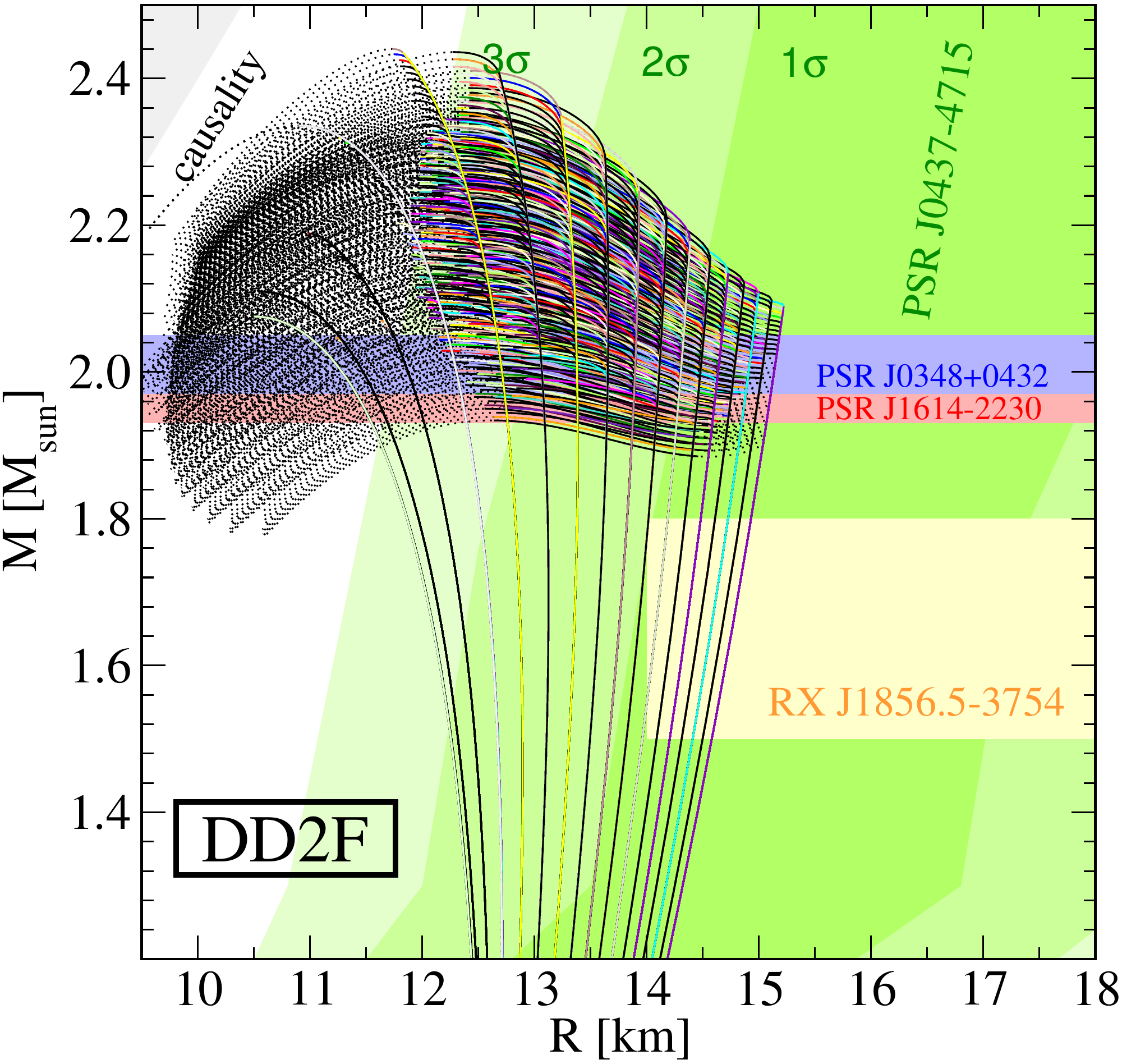} & \includegraphics[width=0.25\textwidth]{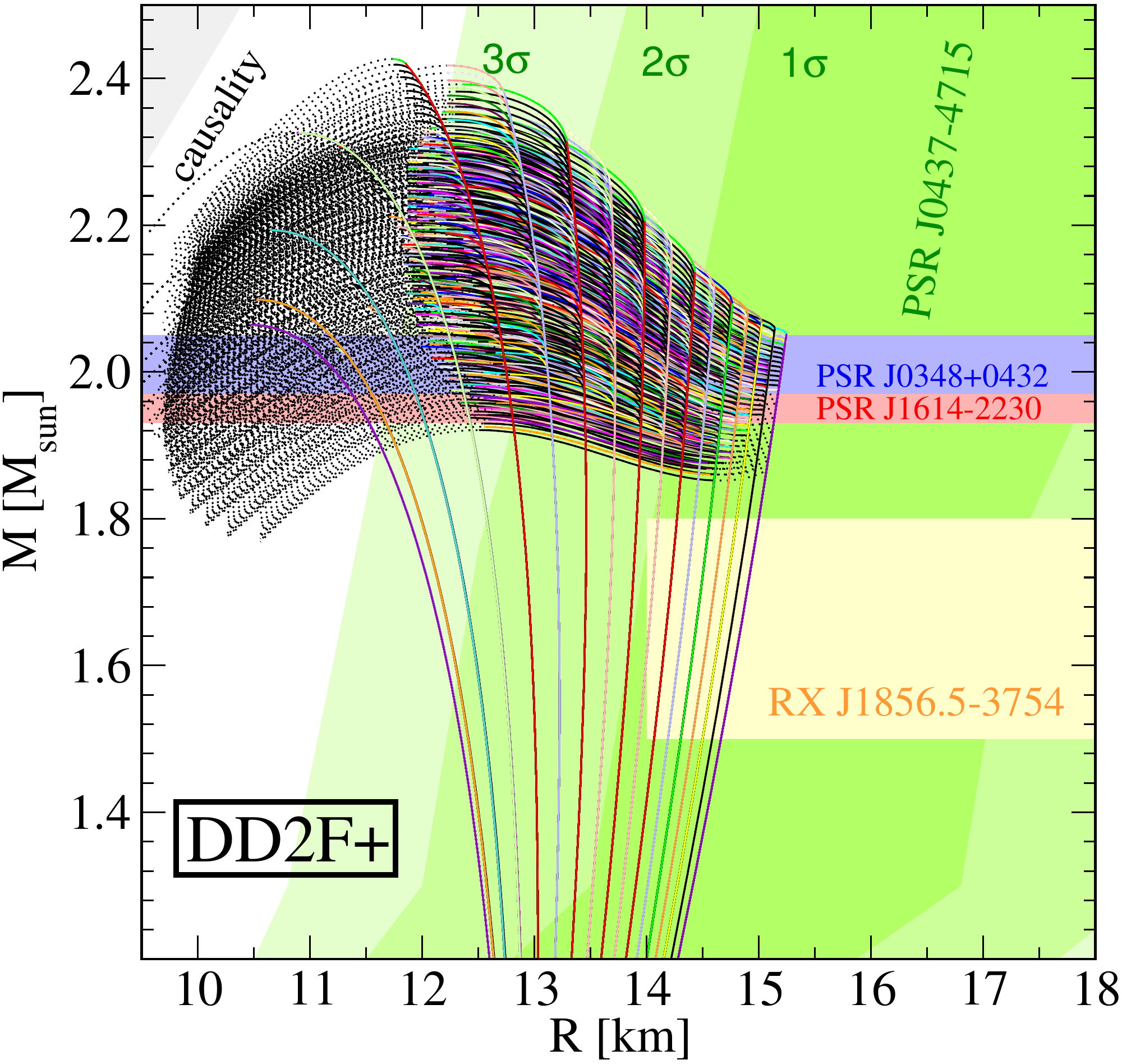}\\
&&&\\
\hline
&&&\\ 
DD2 (stiff)&
\includegraphics[width=0.25\textwidth]{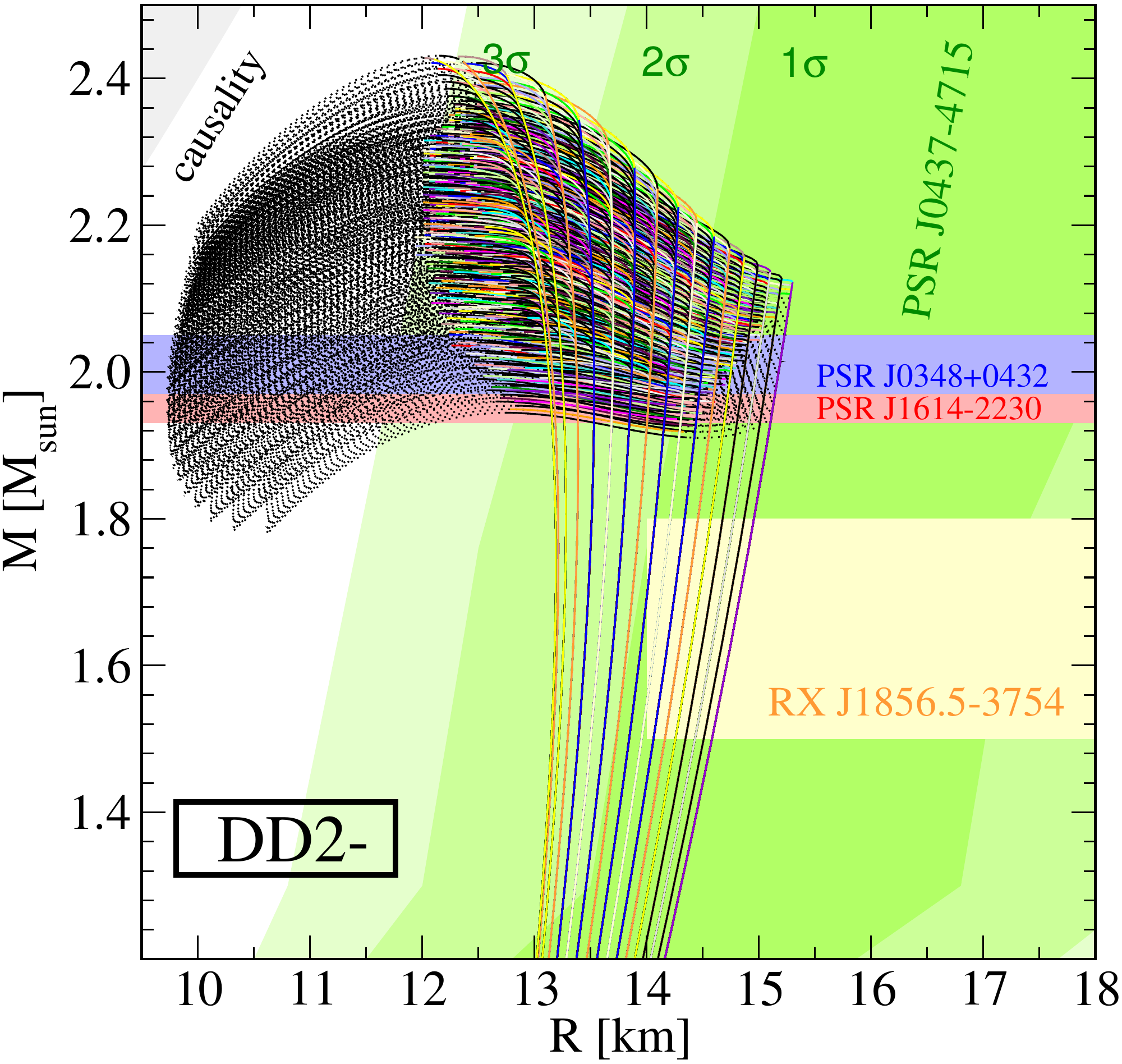} & \includegraphics[width=0.25\textwidth]{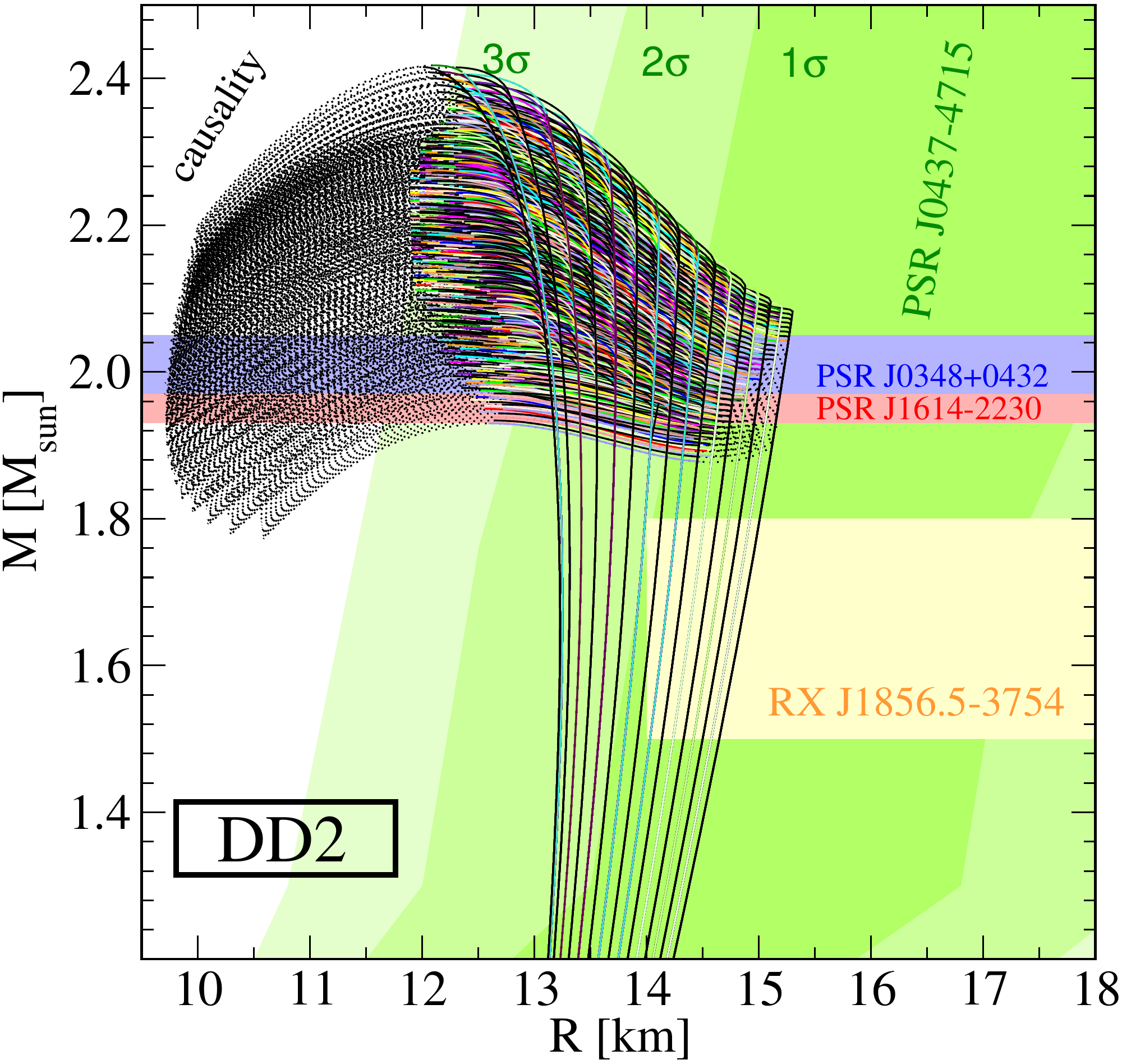} & \includegraphics[width=0.25\textwidth]{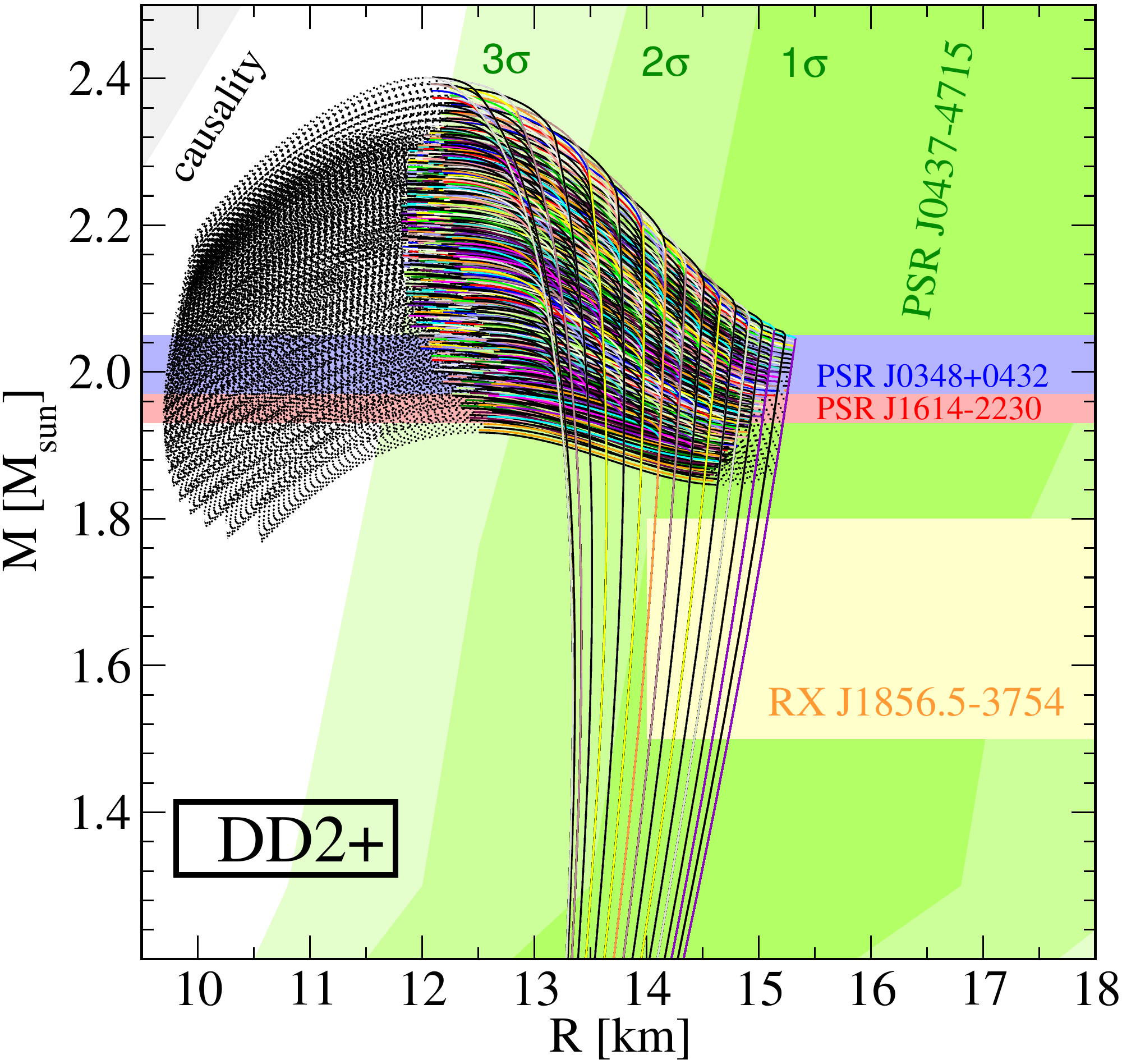}\\
&&&\\
\hline 
\end{tabular}
\end{center} 
\caption{Mass radius relations for the six hybrid EoS classes constructed from the hNJL quark matter EoS and the six hadronic RMF EoS by a Maxwell construction.}
\label{All_M-R}
\end{figure*}

From these two constraints (\ref{beta-eq}) and (\ref{charge}) we can find the total pressure of the 
quark matter phase, $P_q + P_e$, as a function of the 
baryon chemical potential $\mu_B = 2\mu_d+\mu_u$.
The total energy density of the system is
\begin{equation}
\varepsilon = - P_q - P_e + n_u \mu_u + \mu_d n_d + n_e \mu_e~. 
\end{equation}

\subsection{Phase transition: Maxwell construction}

In our study we consider that 
compact stars might undergo a first order phase transition to quark matter. The transition point is determined by enforcing the \textit{Gibbs conditions}: both the pressure and
chemical potential should have the same value in both hadronic and quark matter phases. Whether the resulting compact star will be a hybrid or a pure hadronic will
depend on having these conditions fulfilled at densities lower than the central density of the most massive star of each EoS sequence.

A systematic analysis is performed by varying the EoS parameters both in the hadronic and the quark phase as shown in figure~\ref{Case_A_n_B}. 
This has important consequences for the compactness of the neutron star as well as for the possible existence of a disconnected branch of hybrid stars ("third family") in the mass-radius diagram. 

\subsection{Mass-radius relations for hybrid EoS}

Once the equation of state of neutron star matter is defined, the structure and global properties of compact stars are 
obtained by solving the Tolman-Oppenheimer-Volkoff (TOV) equations~\cite{Tolman:1939jz,Oppenheimer:1939ne} 
\begin{eqnarray}
\frac{dP(r)}{dr}&=& - \frac{G M( r)\varepsilon( r)}{r^2}\frac{\left(1+\frac{P( r)}{\varepsilon( r)}\right)
\left(1+ \frac{4\pi r^3 P( r)}{M( r)}\right)}{\left(1-\frac{2GM( r)}{r}\right)},\nonumber\\
\frac{dM( r)}{dr}&=& 4\pi r^2 \varepsilon( r) ,\nonumber\\
\frac{d N_B( r)}{dr}&=& 4\pi r^2 \left(1-\frac{2GM( r)}{r}\right)^{-1/2}n( r)~.
\end{eqnarray}

The necessary boundary conditions to solve these equations are picking up
a central energy density $\varepsilon_c=\varepsilon(r=0)$ and a central pressure $P_c=P( r=0)$ at $r=0$. The integration is carried out from the center of the star up to
the distance $r=R$ where the pressure vanishes $P( r=R)=0$, defining the radius $R$,
the mass $M=M(R)$ and the baryon number $N_B=N_B(R)$ of the star.
By varying the central energy density is possible to obtain a sequence of star configurations for a given EoS, 
which then uniquely corresponds to a mass-radius curve $M(R)$.


\begin{figure*}[htb!]
\begin{center}
\begin{tabular}{l|c|c|c}
\hline
Symmetry energy $\rightarrow$&&&\\
&soft & medium & stiff\\
EoS $\downarrow$&&&\\
\hline
&&&\\ 
DD2F (semi-soft)&
\includegraphics[width=0.25\textwidth]{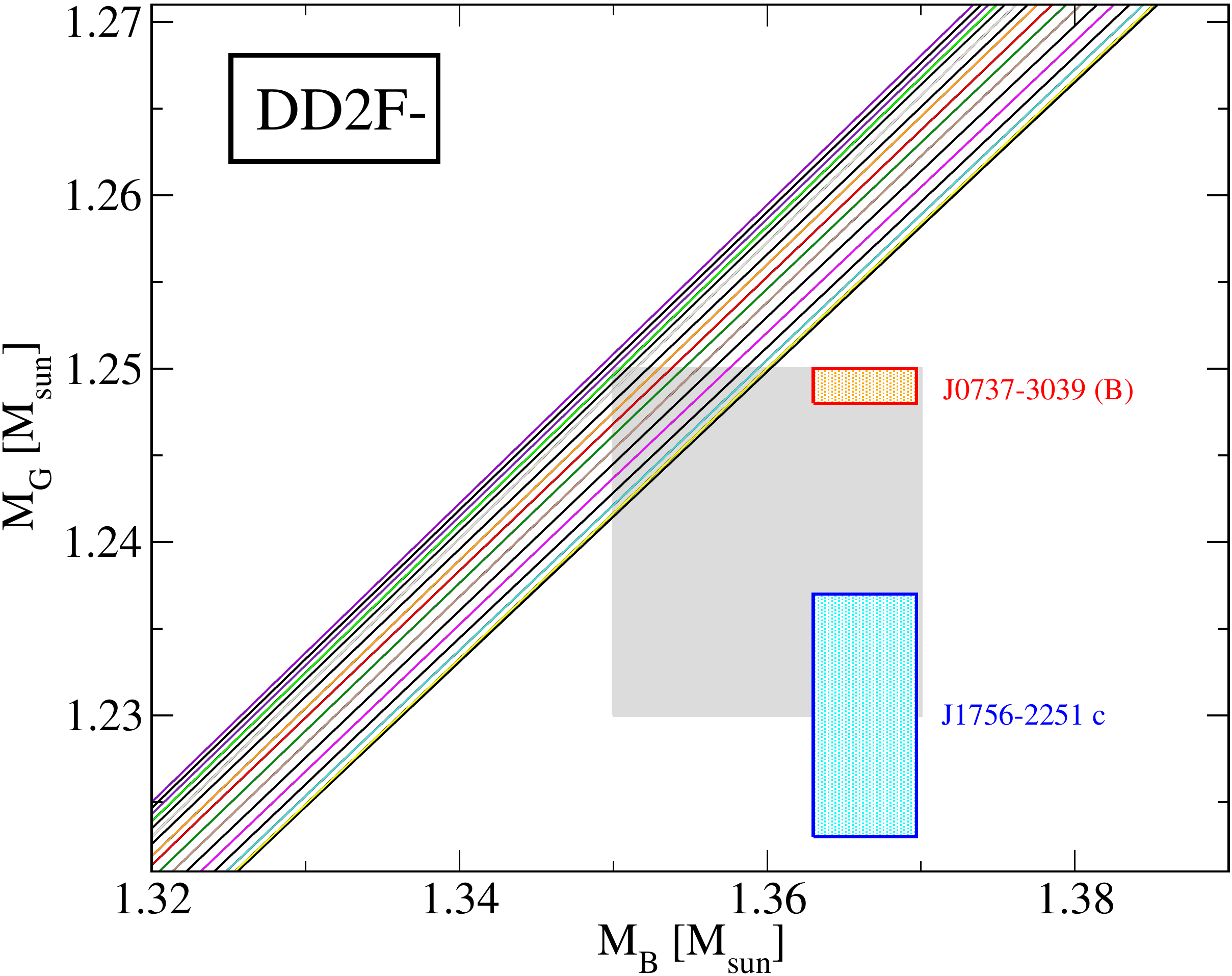} & \includegraphics[width=0.25\textwidth]{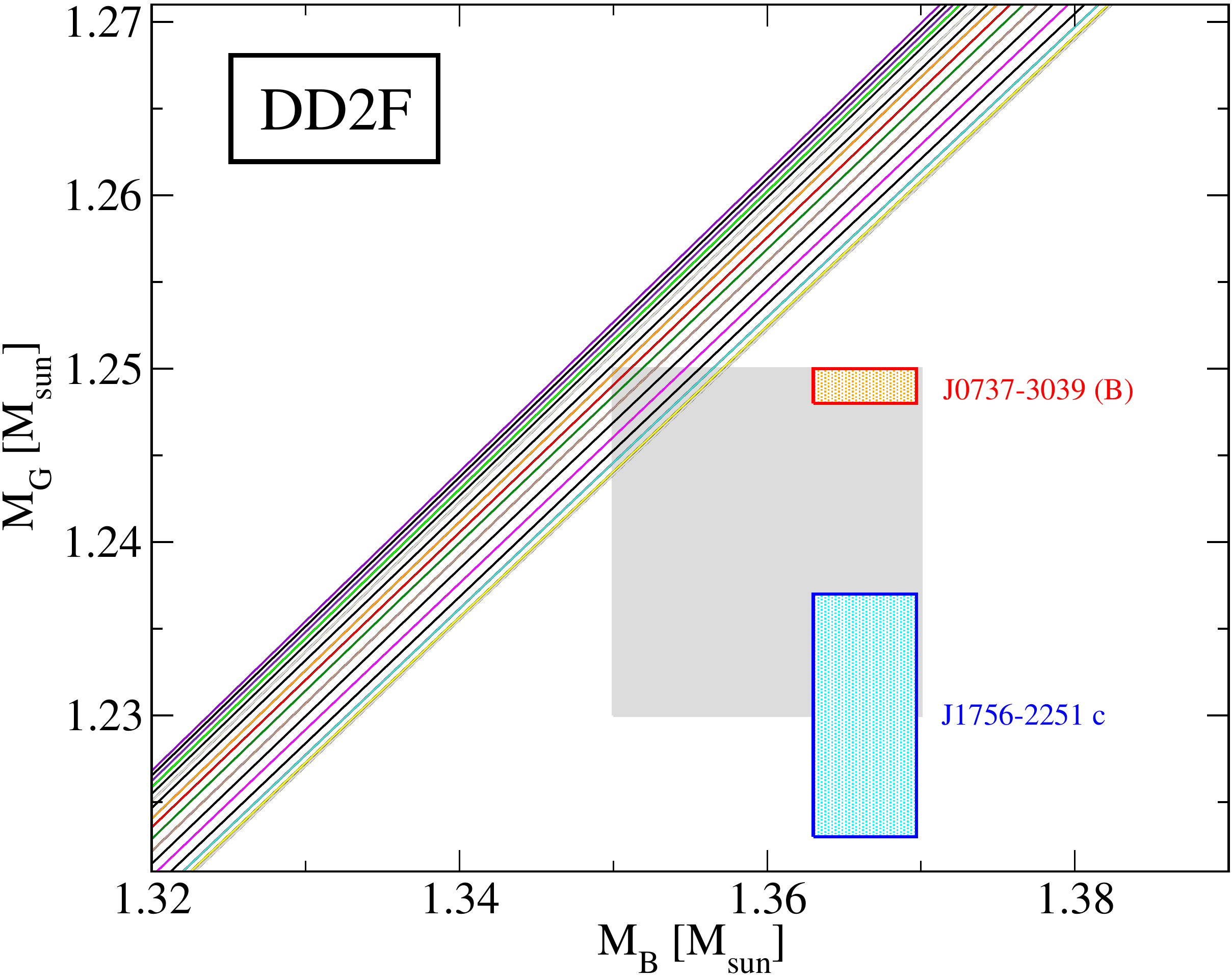} & \includegraphics[width=0.25\textwidth]{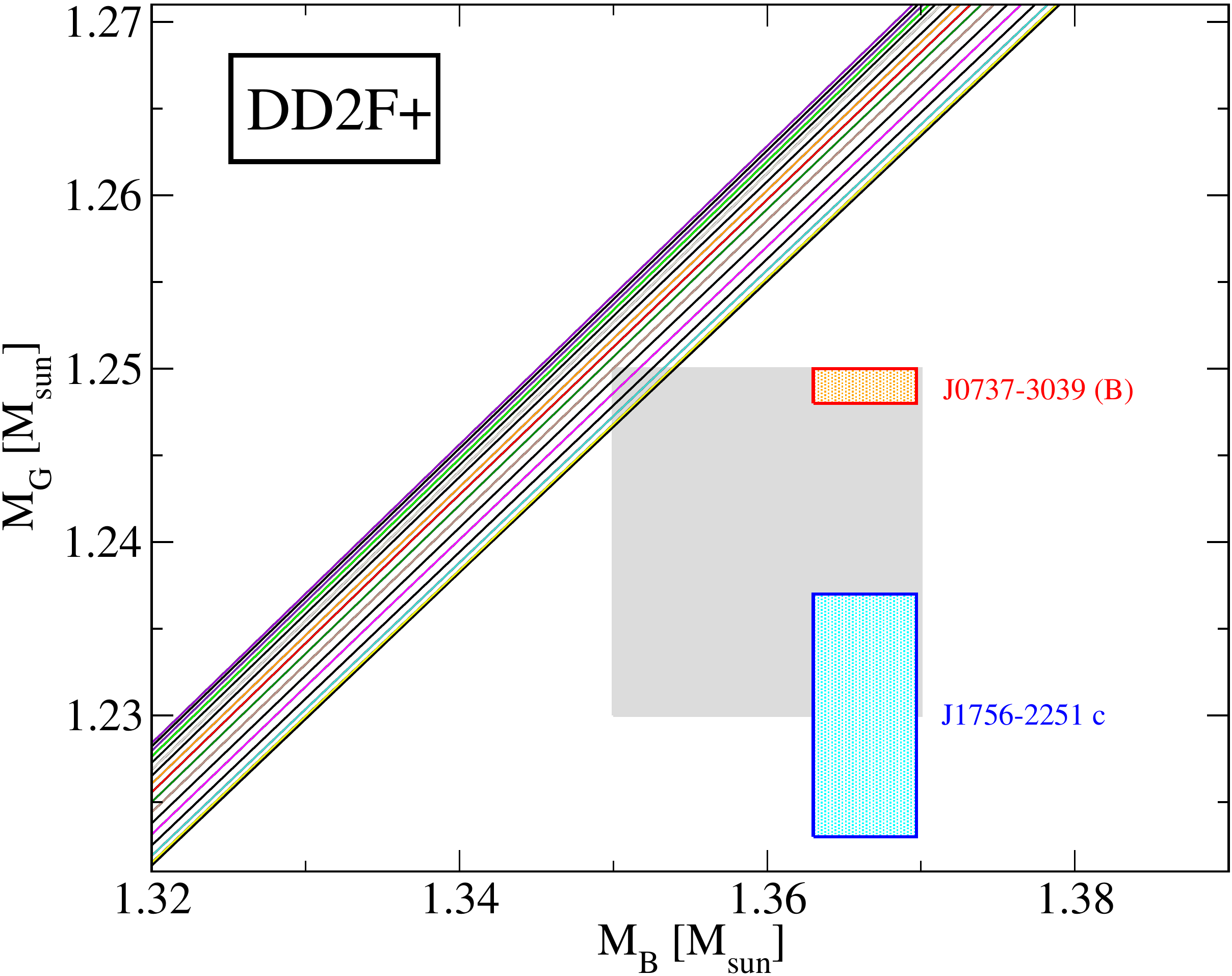}\\
&&&\\
\hline
&&&\\ 
DD2 (stiff)&
\includegraphics[width=0.25\textwidth]{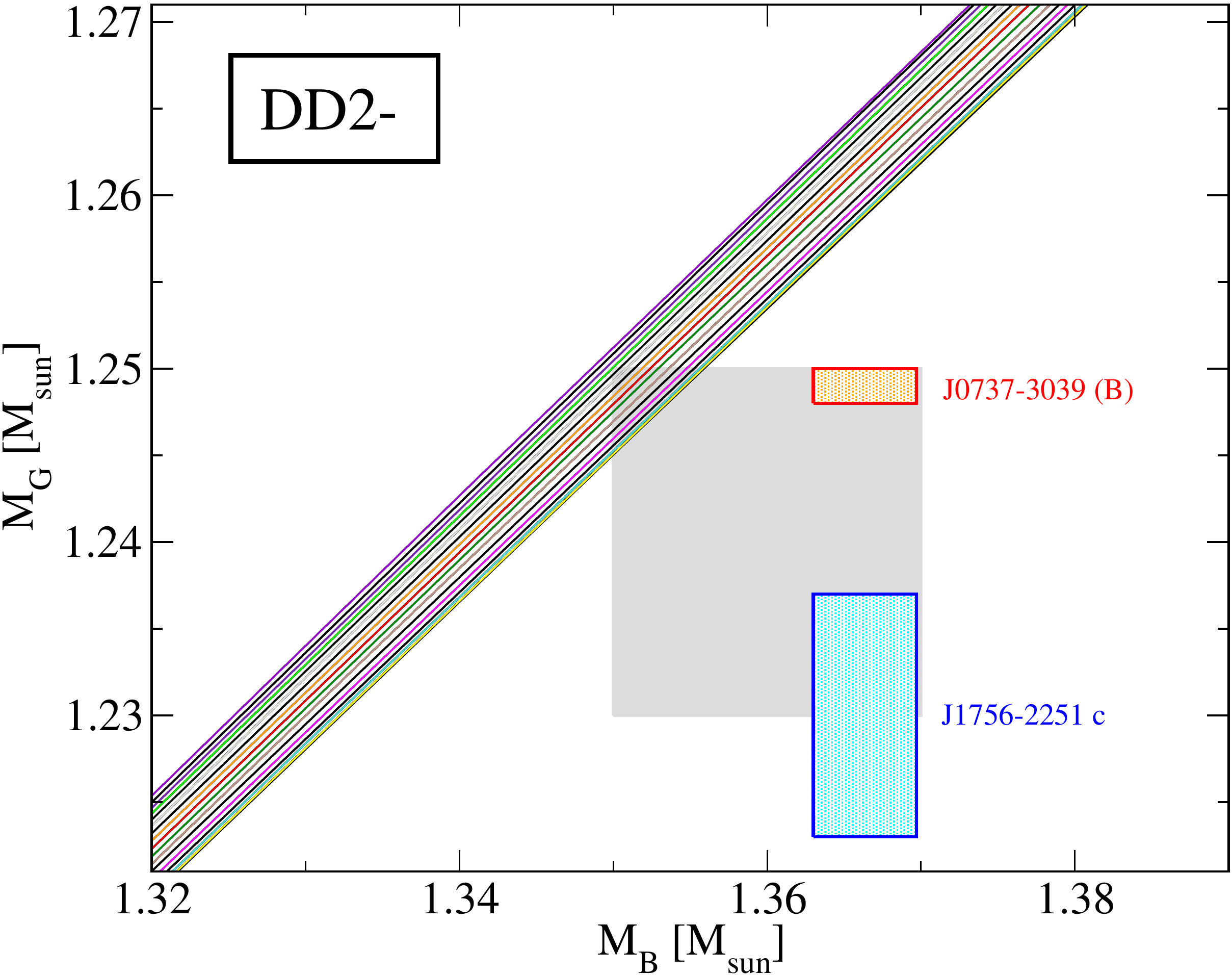} & \includegraphics[width=0.25\textwidth]{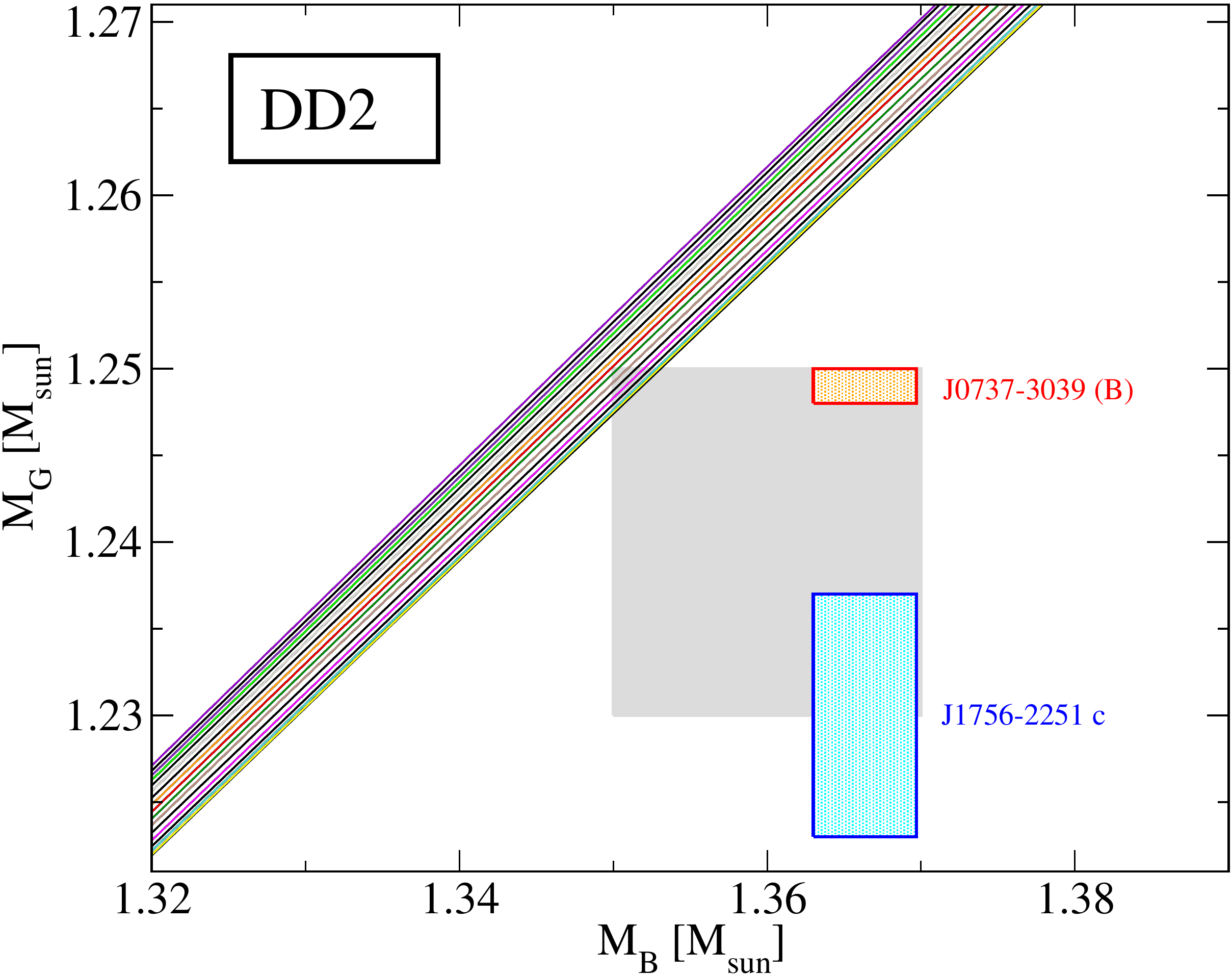} & \includegraphics[width=0.25\textwidth]{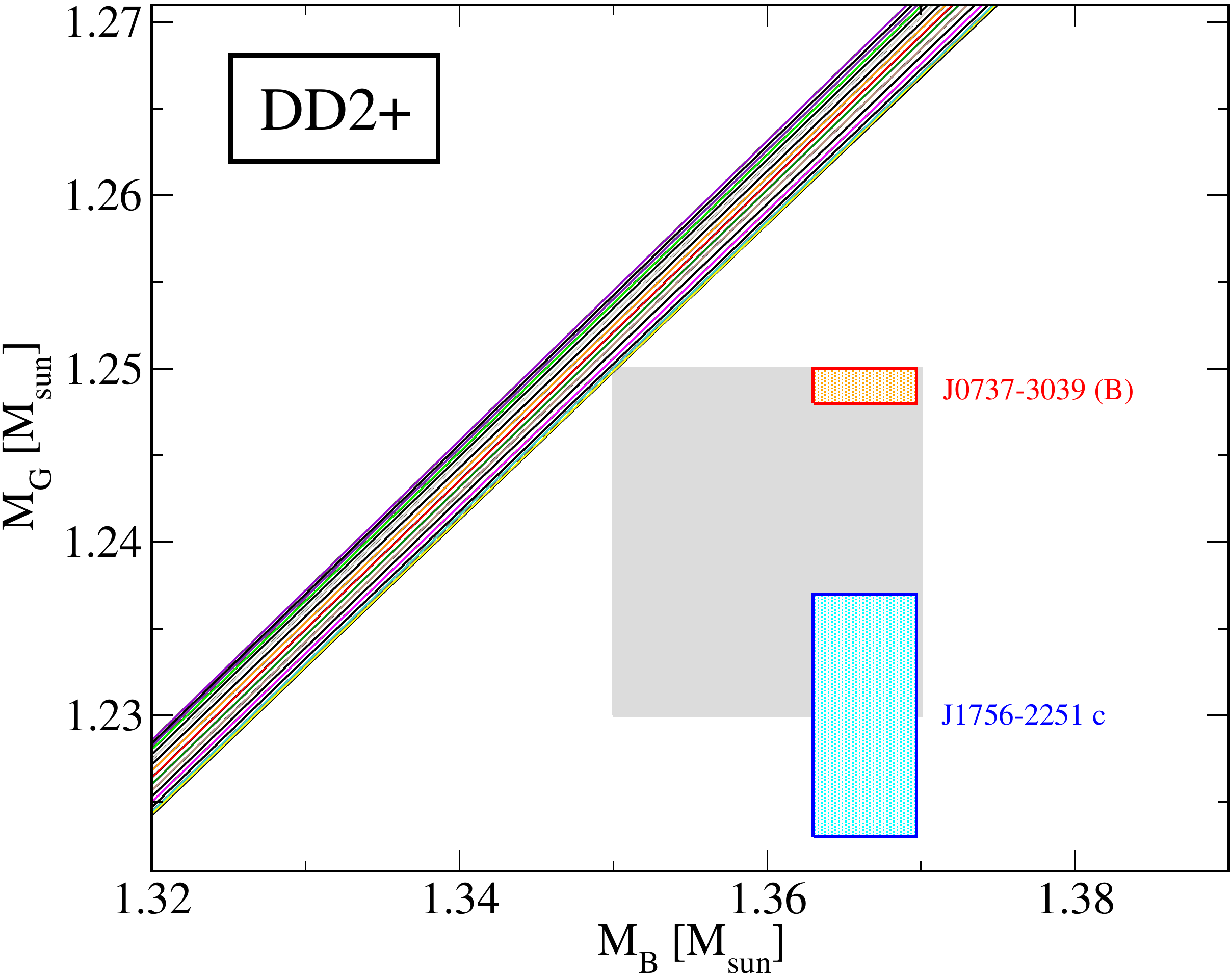}\\
&&&\\
\hline 
\end{tabular}
\end{center} 
\caption{Gravitational mass vs. baryonic mass for the six classes of new hybrid EoS.
For the explanation of the boxes, see text.
}
\label{All_Mg-Mb}
\end{figure*}

\section{Bayesian analysis for Compact Stars}
\label{sec:bayesian}

Bayesian methods have proven to be a poweful technique for model descrimination and parameter descrimination.
In a primer Bayesian study performed by Steiner et al. \cite{Steiner:2010fz} the luminosity of expanding photospheric radius 
extracted for burst sources has been used to constrain a combined mass-radius 
relationship. However, this method is problematic, in particular because of the unknown stellar 
atmosphere composition, uncertainties in the distance to the 
source, the bias of the parabolic M-R constraint with the shape of stellar
sequences in the M-R diagram for typical EoS and last but not least due to unknown 
details of the burst mechanism. In this work we do BA as well but we implement different observables which will be used as constraints:
the highest measured mass~\cite{Antoniadis:2013pzd}, a radius
measurement~\cite{Bogdanov:2012md} and a gravitational mass vs. baryonic mass determination by
Podsiadlowski et al. \cite{Podsiadlowski:2005ig}. 
These constraints were already included in our earlier Bayesian analyses, e.g., of 
Refs.~\cite{Alvarez-Castillo:2014xea,Blaschke:2014via,Ayriyan:2015kit}. 
We will describe the Bayesian method and the chosen constraints in the following subsections.

\subsection{Bayesian analysis technique}

We start by defining a vector of free parameters $\overrightarrow{\pi}=\{p,\eta_{4}\}$,
which correspond to all the possible models with or without phase transition
from nuclear to quark matter using the EoS described above. The way
we sample these parameters is
\begin{equation}
\overrightarrow{\pi}_{i}=\left\{ p_{(k)},\eta_{4(l)}\right\} ,\label{pi_vec}
\end{equation}
where $i=0\dots N-1$ with $N=N_{1}\times N_{2}$ such that $i=N_{2}\times k+l$
and $k=0\dots N_{1}-1$, $l=0\dots N_{2}-1$, with $N_{1}$ and $N_{2}$
being the total number of parameters $p_{(k)}$ and $\eta_{4(l)}$,
respectively.
After integration of the TOV equation each EoS model will provide estimations for NS properties that we shall use for Bayesian analysis. 
Thus, these results allow us to use different neutron star observations in order to determine the
probability that a given EoS fulfils the observational constraints.
Our goal is to find the set of most probable $\overrightarrow{\pi}_{i}$
matching the above constraints using the BA technique. For initializing
the BA we propose that \textit{a priori} each vector of parameters
$\overrightarrow{\pi}_{i}$ have the same probability, $P\left(\overrightarrow{\pi}_{i}\right)=1/N$,
for all $i$.

\begin{figure*}[!t]
\begin{center}
\begin{tabular}{l|c|c|c}
\hline
Symmetry energy $\rightarrow$&&&\\[-2mm]
&soft & medium & stiff\\[-2mm]
EoS $\downarrow$&&&\\
\hline
&&&\\[-2mm] 
DD2F (semi-soft)&
\includegraphics[width=0.25\textwidth]{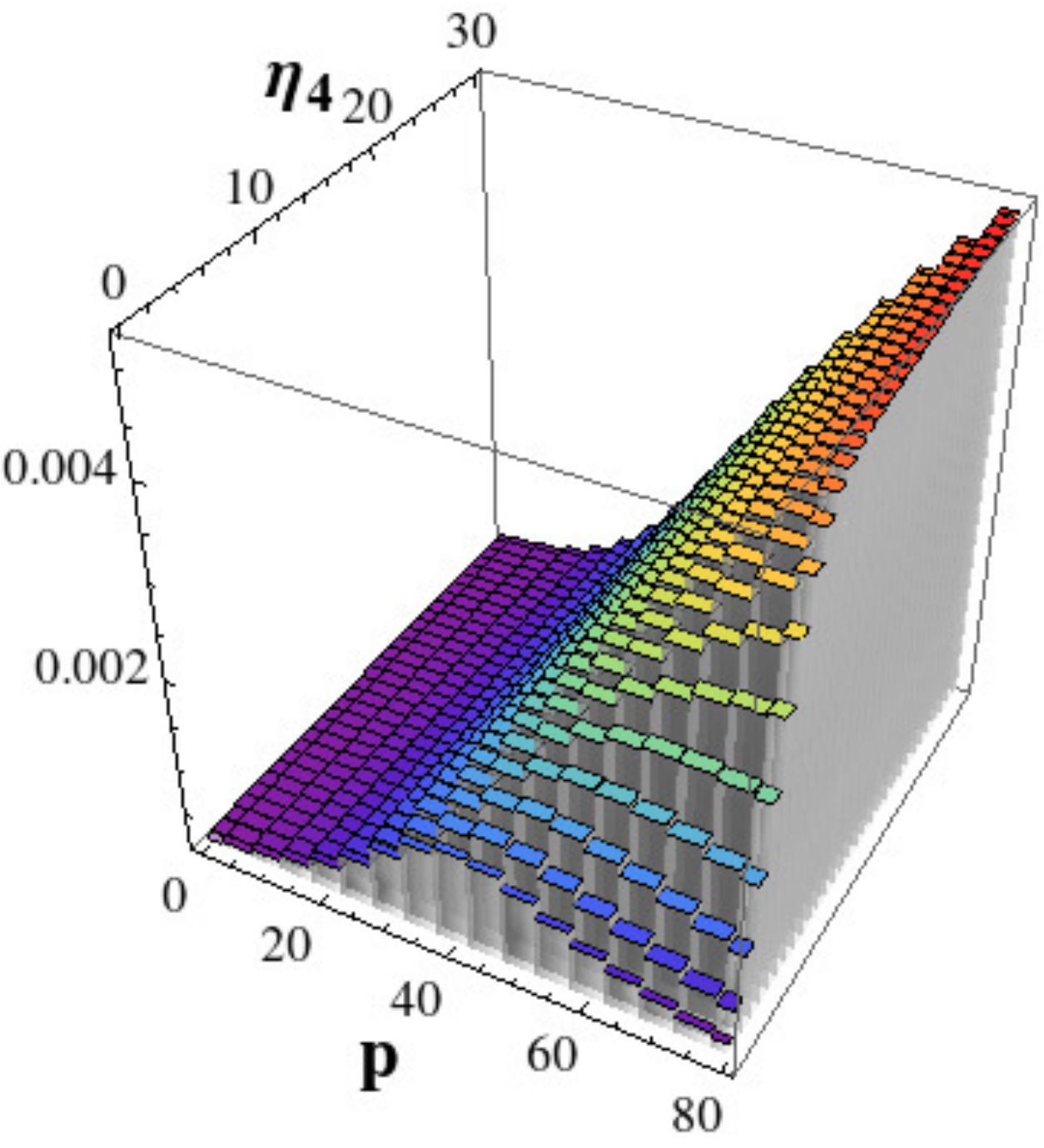} & \includegraphics[width=0.25\textwidth]{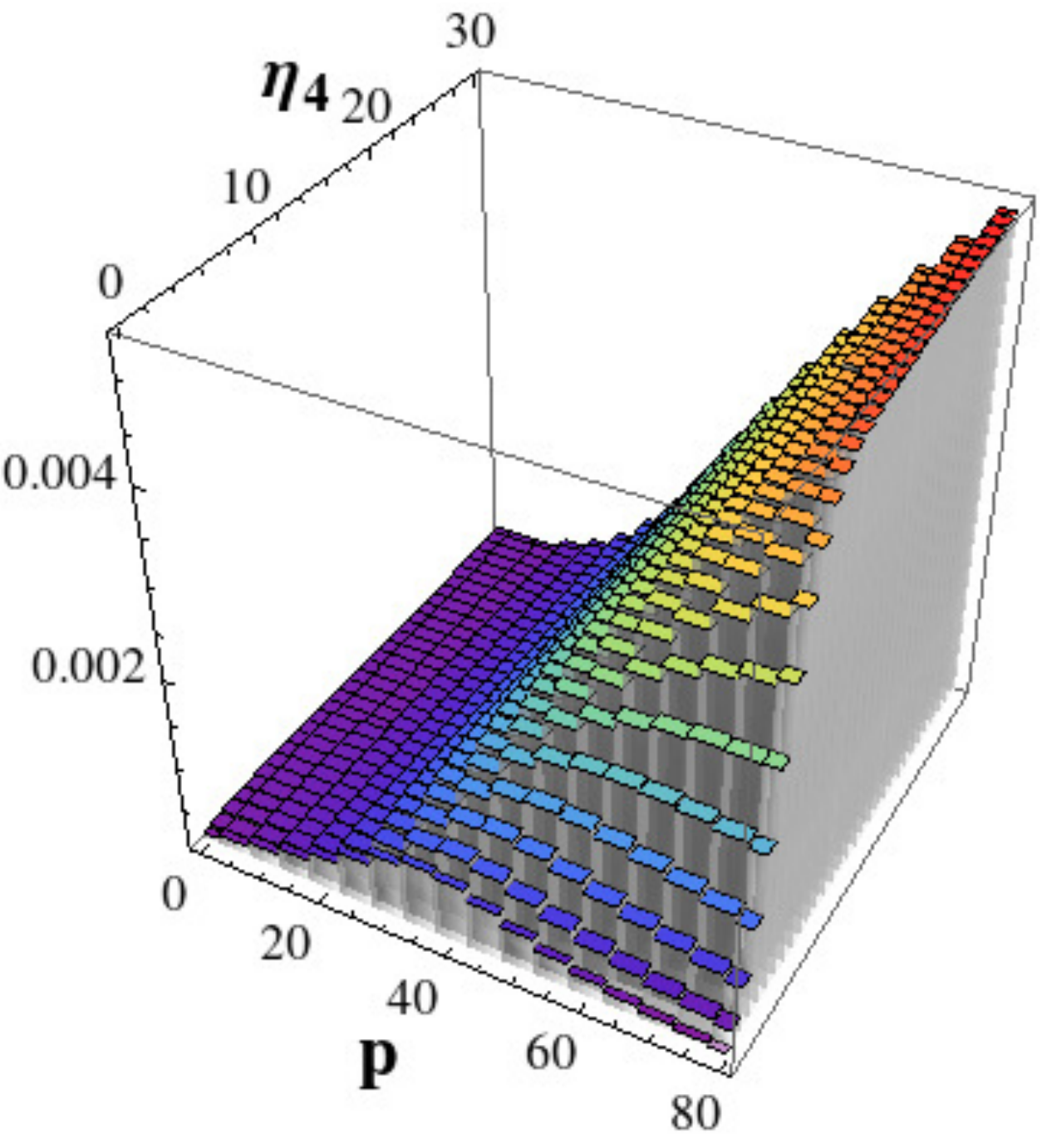} & \includegraphics[width=0.25\textwidth]{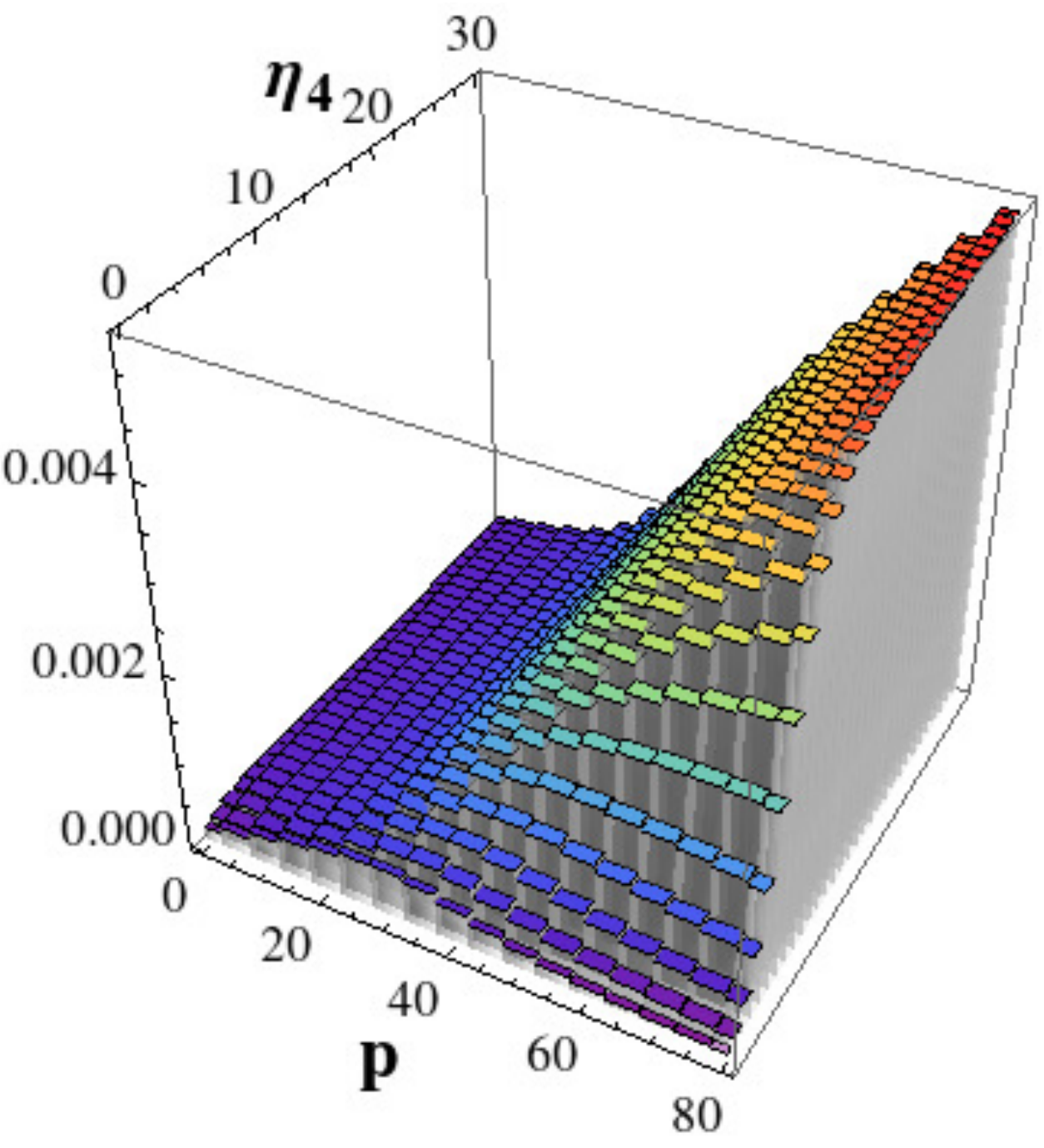}\\
&&&\\[-2mm]
\hline
&&&\\[-2mm] 
DD2 (stiff)&
\includegraphics[width=0.25\textwidth]{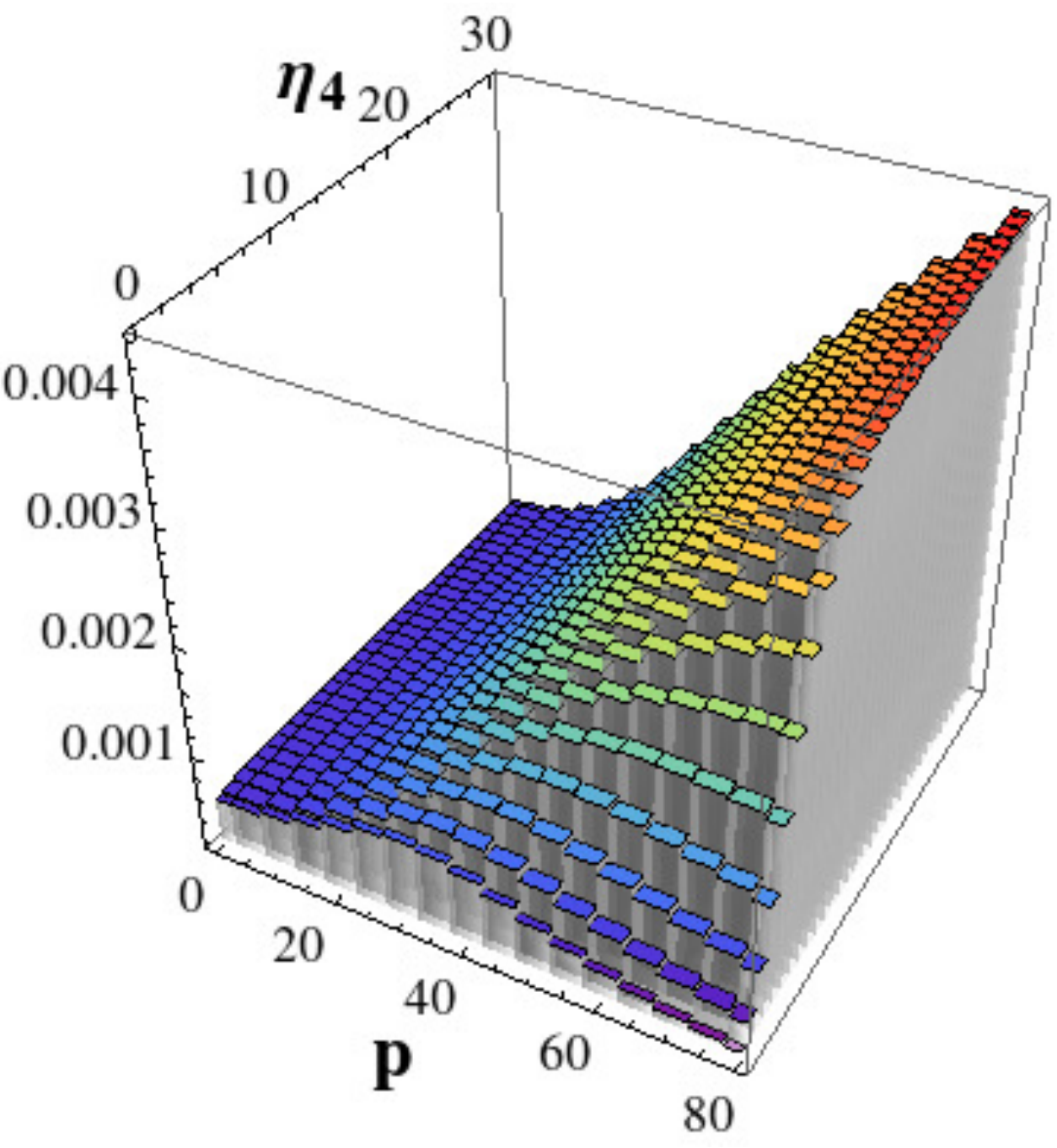} & \includegraphics[width=0.25\textwidth]{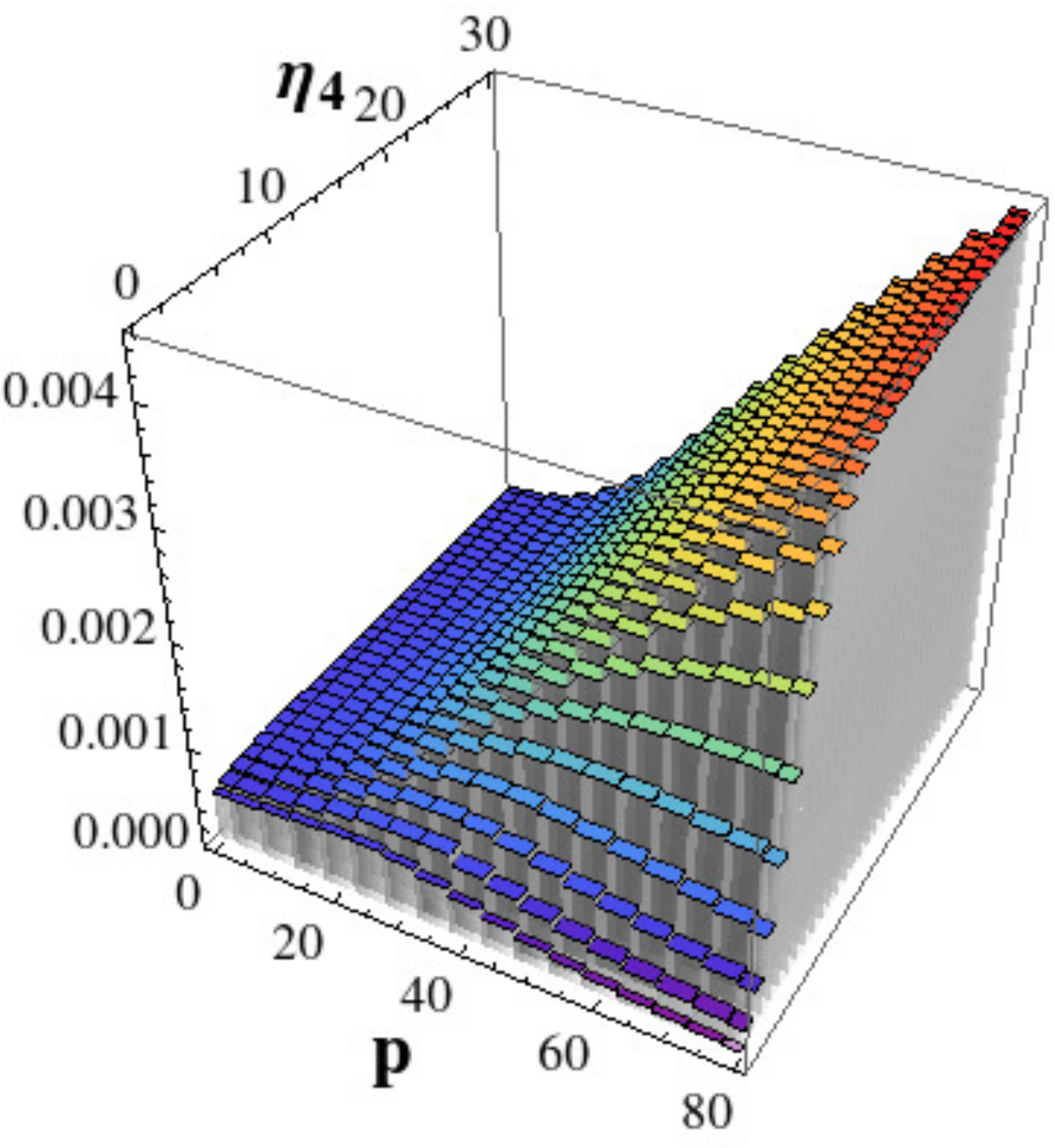} & \includegraphics[width=0.25\textwidth]{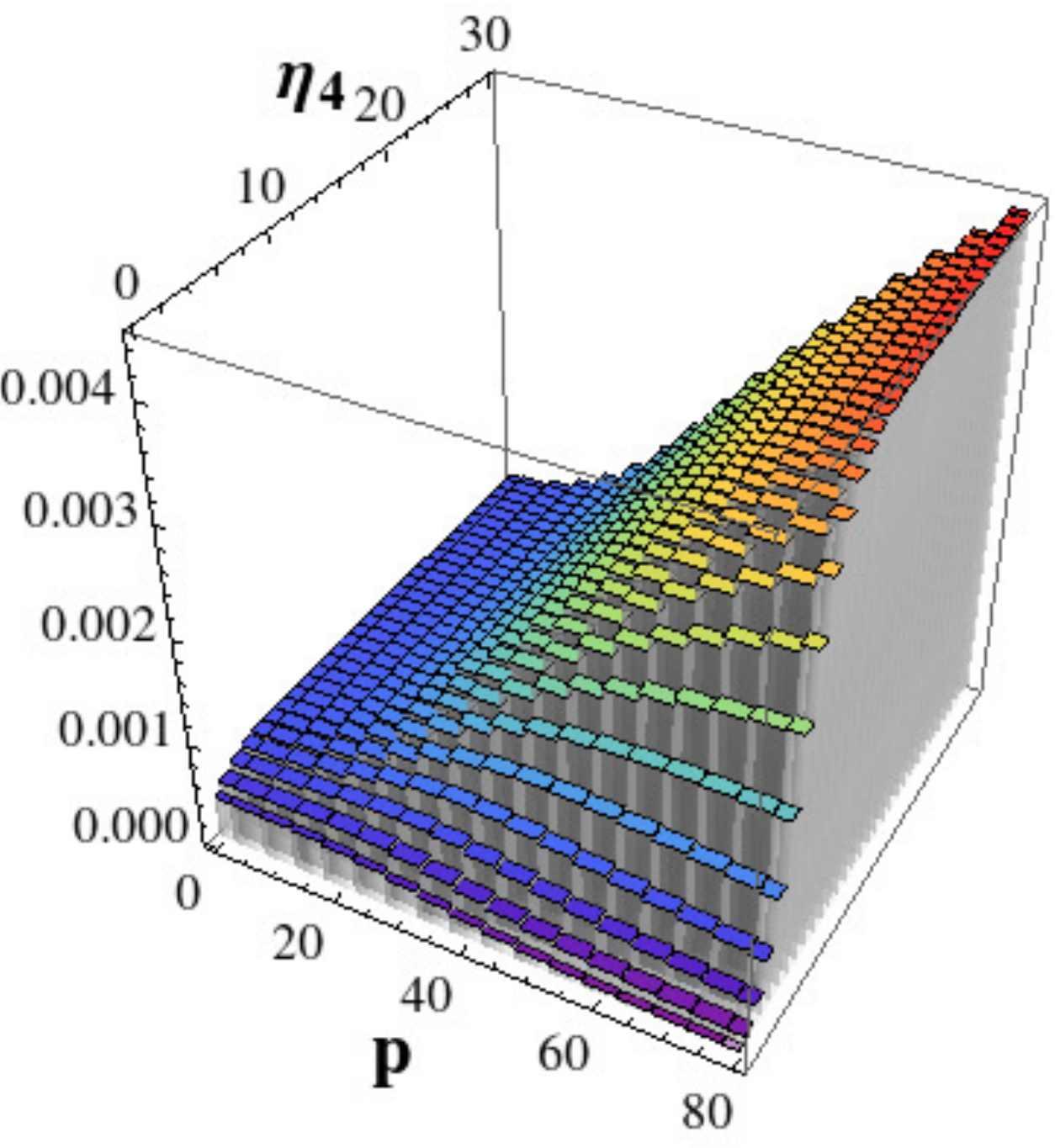}\\
&&&\\
\hline 
\end{tabular}
\end{center}
\caption{Probabilities without the gravitational mass vs. baryon mass constraint.
The stiffest possible EoS are favored, corresponding to large radii and small gravitational binding.}
\label{First_Lego}
\end{figure*}

\begin{figure*}[!htb]
\begin{center}
\begin{tabular}{l|c|c|c}
\hline
Symmetry energy $\rightarrow$&&&\\[-2mm]
&soft & medium & stiff\\[-2mm]
EoS $\downarrow$&&&\\
\hline
&&&\\[-2mm] 
DD2F (semi-soft)&
\includegraphics[width=0.25\textwidth]{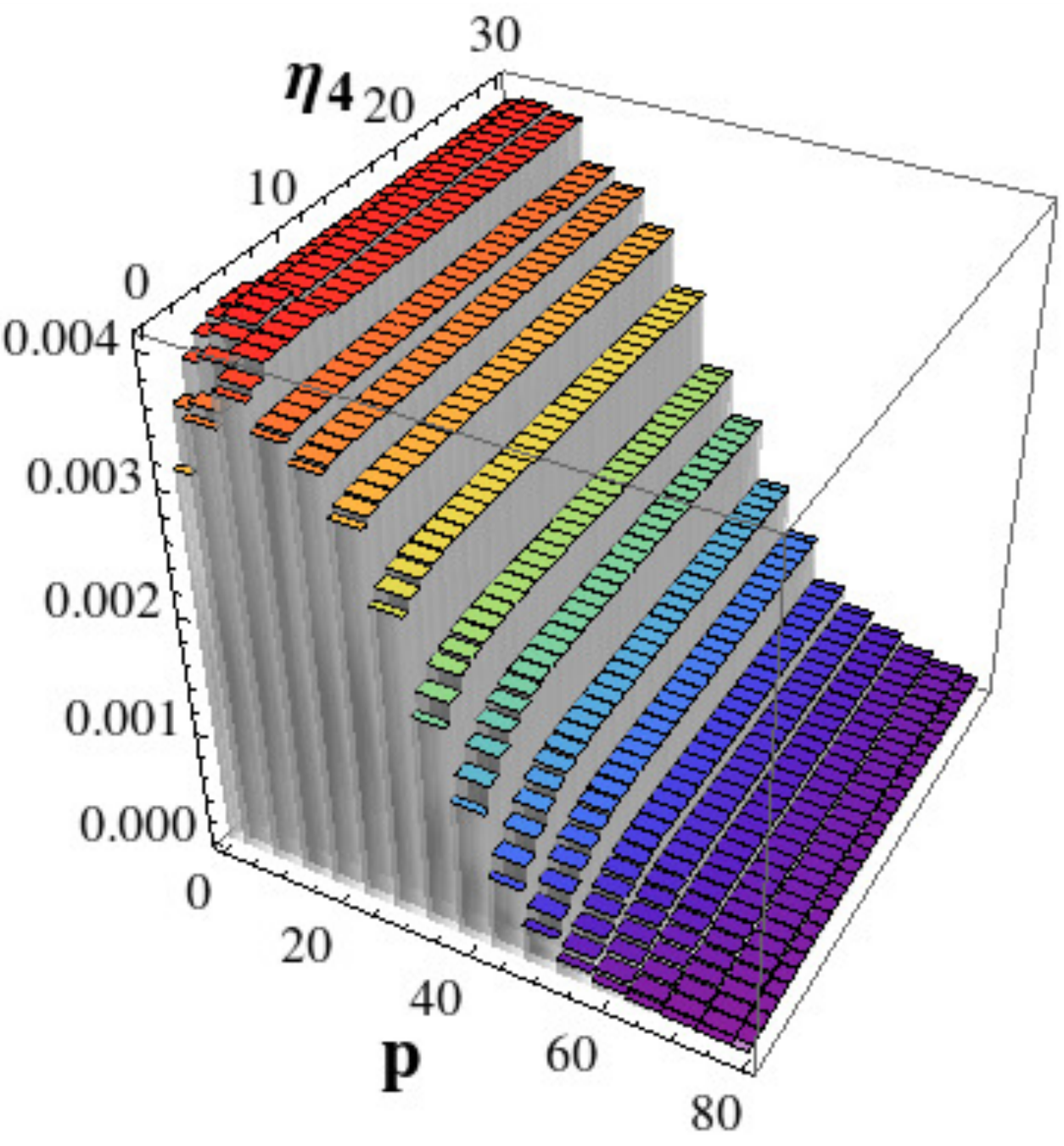} & \includegraphics[width=0.25\textwidth]{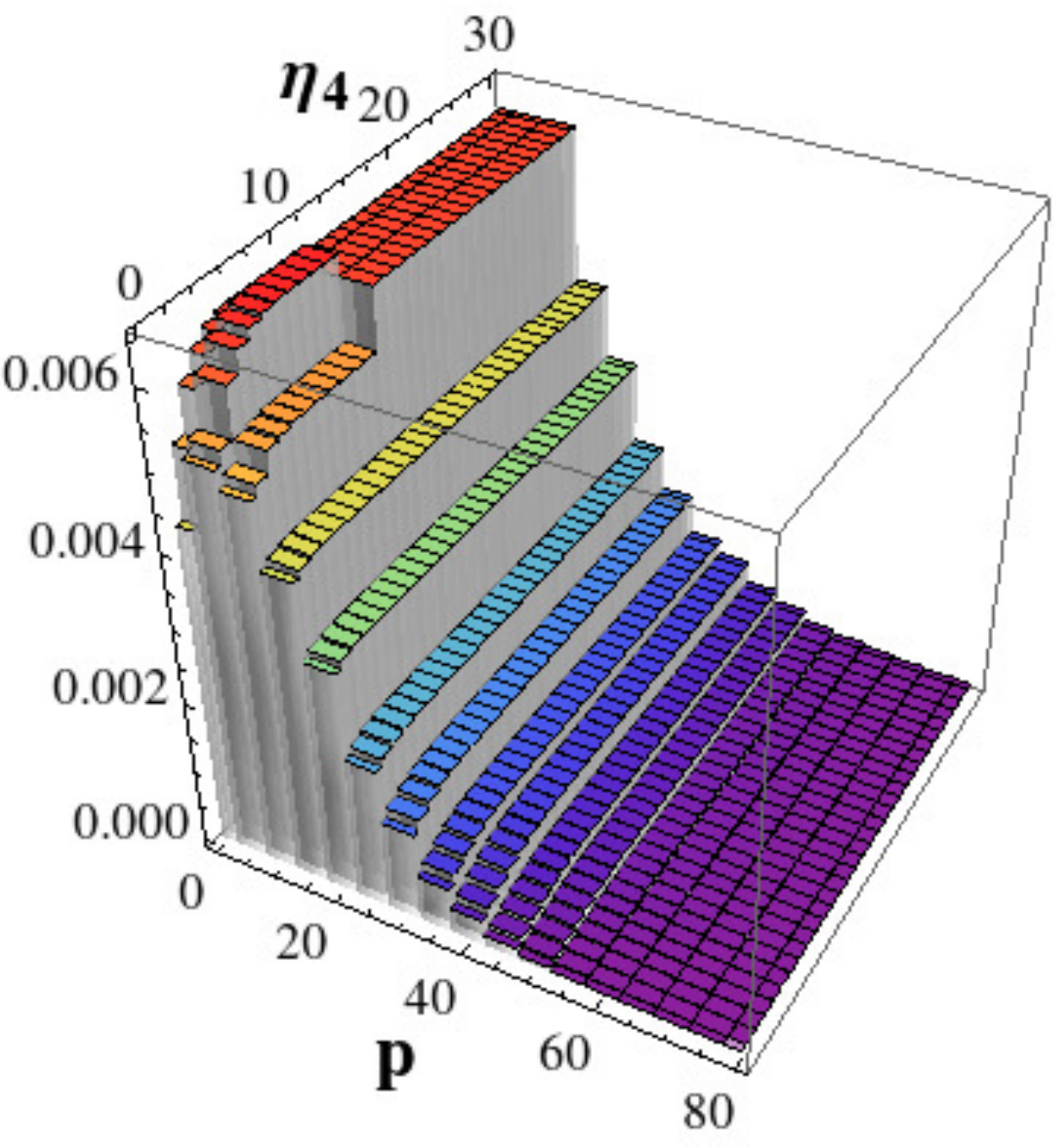} & \includegraphics[width=0.25\textwidth]{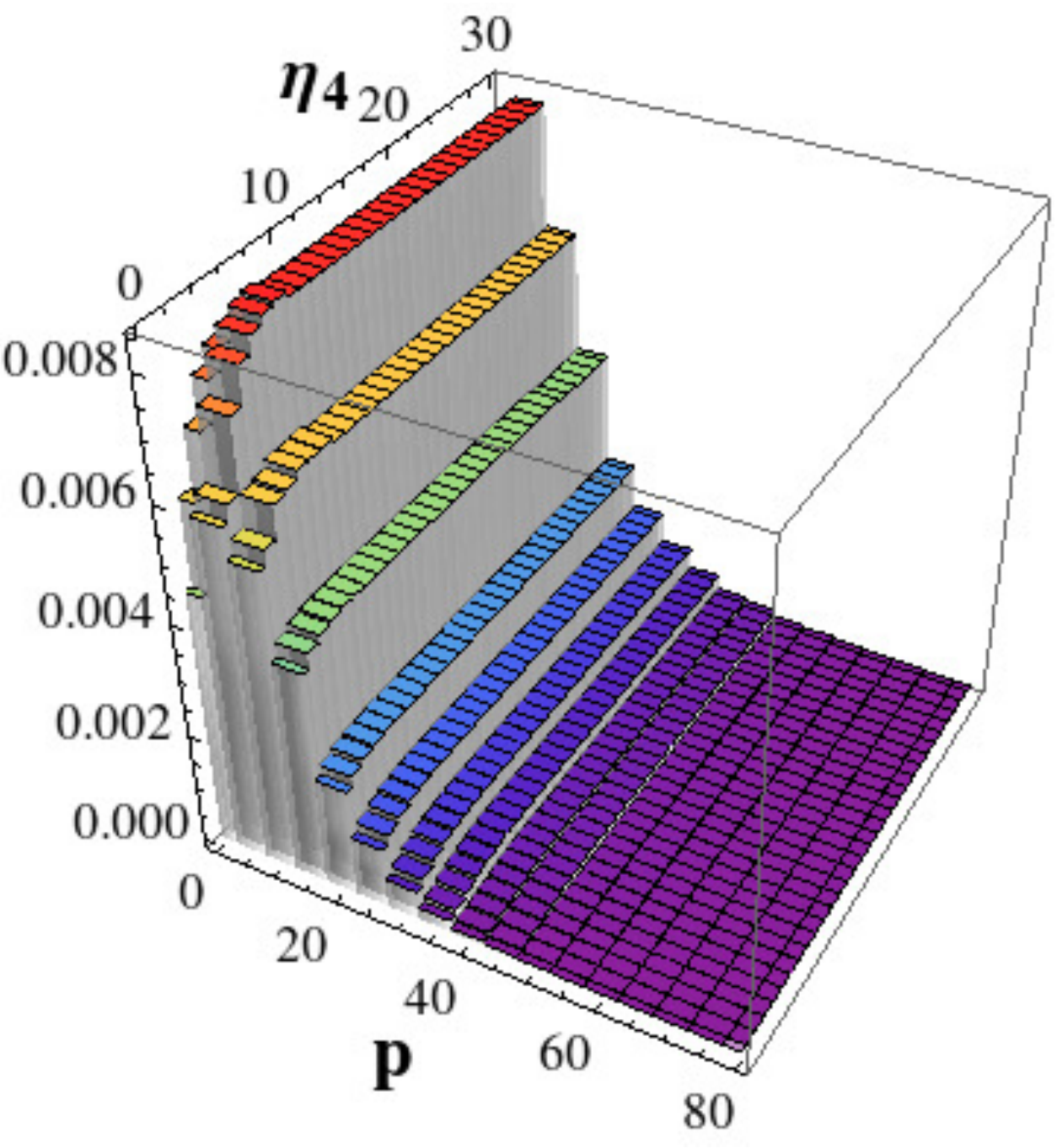}\\
&&&\\[-2mm]
\hline
&&&\\[-2mm] 
DD2 (stiff)&
\includegraphics[width=0.25\textwidth]{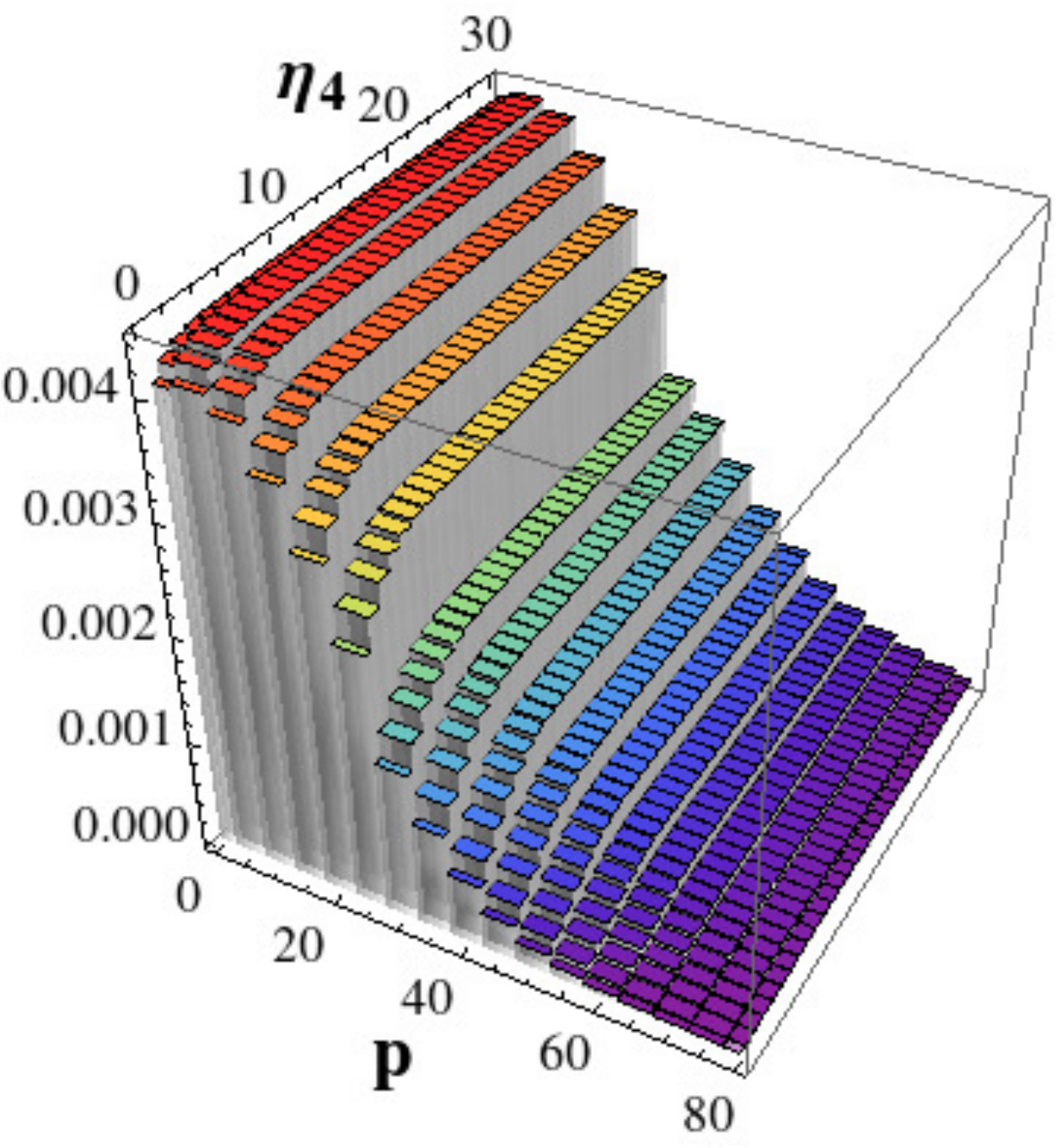} & \includegraphics[width=0.25\textwidth]{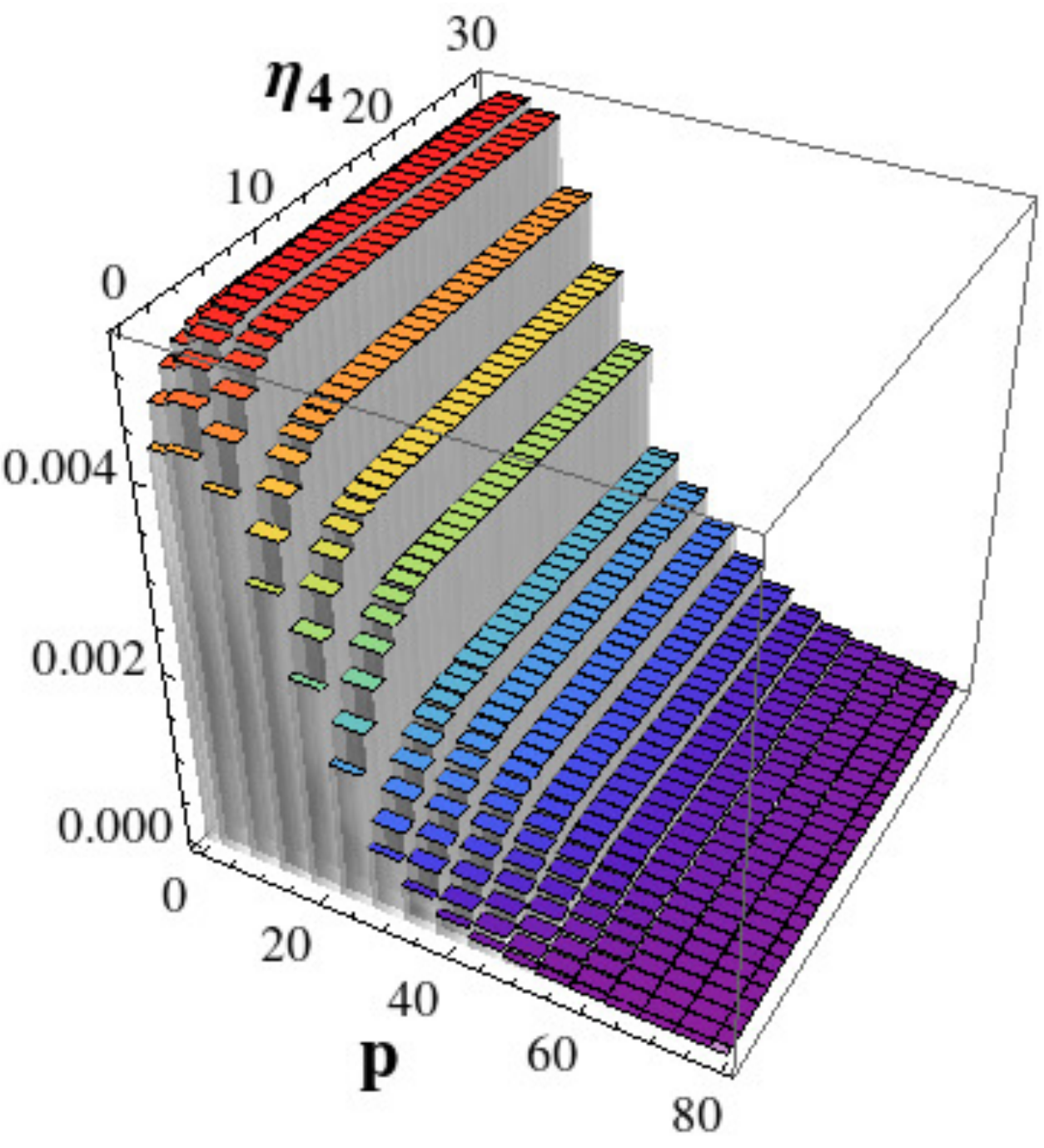} & \includegraphics[width=0.25\textwidth]{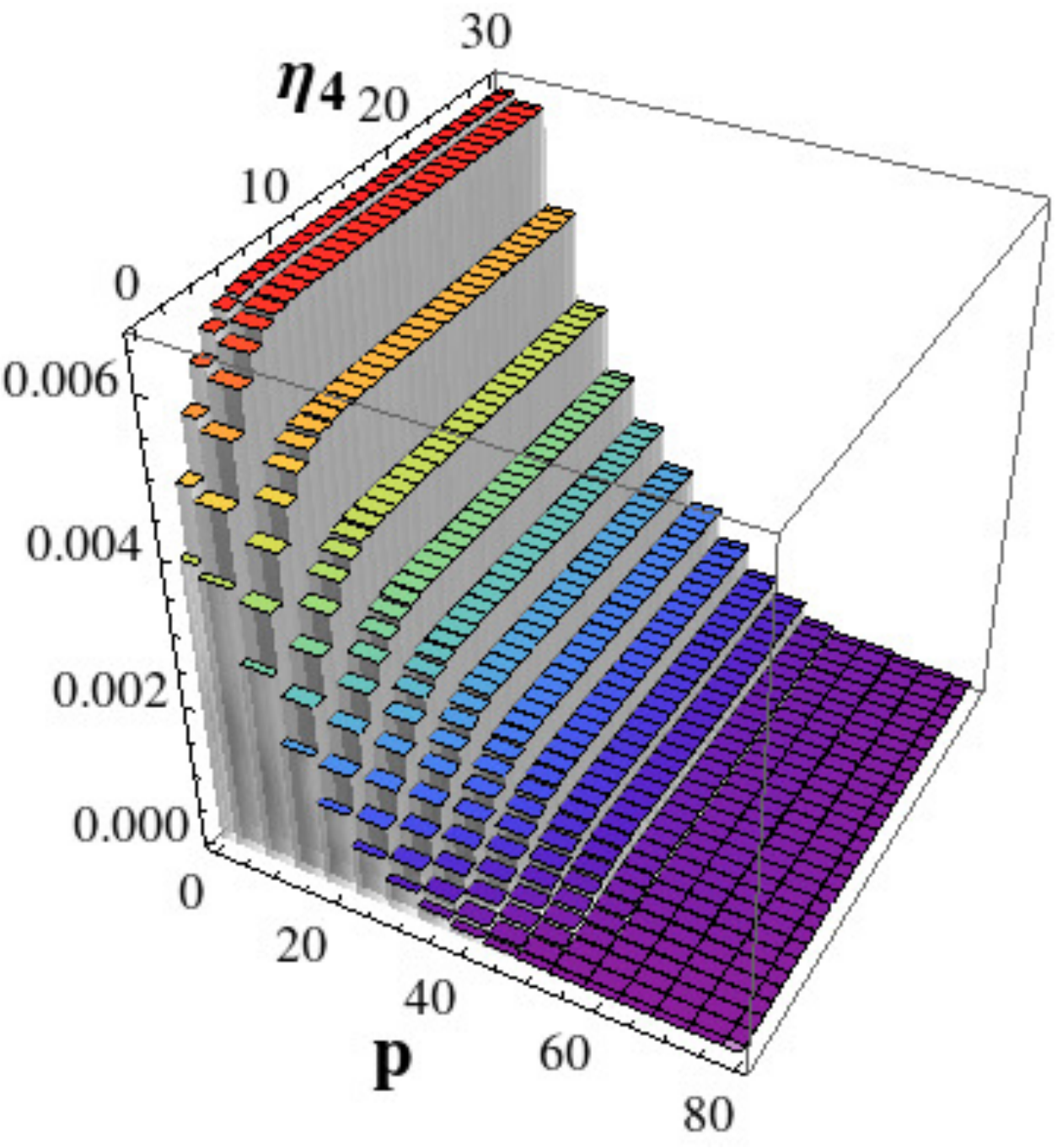}\\
&&&\\
\hline 
\end{tabular}
\end{center} 
\caption{Probabilities with the gravitational mass vs. baryon mass constraint.
The picture of Fig.~\ref{First_Lego} is reversed: now the soft EoS with small excluded volume parameter $p$
are favored which correspond to stars with small radii and large gravitational binding.}
\label{Second_Lego}
\end{figure*}

\subsection{Mass constraint}

We describe the error for the event $E_{A}$ of a mass measurement
of the high-mass pulsar PSR~J0348+0432~\cite{Antoniadis:2013pzd}
with a normal distribution $\mathcal{N}(\mu_{A},\sigma_{A}^{2})$,
where the mean value of the mass is $\mu_{A}=2.01~\mathrm{M_{\odot}}$
and the variance is $\sigma_{A}=0.04~\mathrm{M_{\odot}}$. Using this
assumption we compute the conditional probability of the event $E_{A}$
(under the condition that the neutron star is described by the EoS
model with the parameters $\overrightarrow{\pi}_{i}$) with
\begin{equation}
P\left(E_{A}\left|\overrightarrow{\pi}_{i}\right.\right)=\Phi(M_{i},\mu_{A},\sigma_{A})~.\label{p_anton}
\end{equation}
Here $M_{i}$ is the maximum mass accessible with the vector $\overrightarrow{\pi}_{i}$
and $\Phi(x,\mu,\sigma)$ is the cumulative distribution function
for the Gaussian distribution
\begin{equation}
\Phi(x,\mu,\sigma)=\frac{1}{2}\left[1+{\rm erf}\left(\frac{x-\mu}{\sqrt{2\sigma^{2}}}\right)\right].\label{Laplas}
\end{equation}

\subsection{Radius constraint}

We consider here a very promising technique to measure radii of neutron
stars that is based on the pulse phase resolved X-ray spectroscopy
which properly accounts for the system geometry of a radio pulsar.
This radius measurement gives $\mu_{B}=15.5~\mathrm{km}$ and $\sigma_{B}=1.5~\mathrm{km}$
for PSR~J0437-4715~\cite{Bogdanov:2012md}. 
Moreover, Hambaryan et al. \cite{Hambaryan:2014} have also reported compatible radius
measurements for RXJ 1856.5-3754.

We compute the conditional probability of the event $E_{B}$ that
the measured radius of the neutron star corresponds to the model with
$\overrightarrow{\pi}_{i}$ as
\begin{equation}
P\left(E_{B}\left|\overrightarrow{\pi}_{i}\right.\right)=\Phi(R_{i},\mu_{B},\sigma_{B})~.\label{p_bogdan}
\end{equation}
Here the value $R_{i}$ is the maximum radius for the given vector
$\overrightarrow{\pi}_{i}$.

\subsection{Gravitational mass vs. baryonic mass constraint}

For the selection of the EoS with different symmetry energies we
consider as an observational constraint the properties of the star B in the double pulsar 
PSR~J0737-3039(B) \cite{Kramer:2006nb} 
Following \cite{Podsiadlowski:2005ig}, 
the evolution of this object based on the hypothesis that the neutron
star is born in a supernova collapse after electron capture instability of an O-Ne-Mg star. 
This method allows us to estimate the mass of the progenitor core, being
approximately the baryon mass of the new-born neutron star.
The measured neutron star gravitational mass is $M_{G}=1.249~M_{\odot}$, with error-bar $\Delta M=\pm 0.001~M_{\odot}$, while from the evolution model the baryon mass is expected to be $\mu_{B}=1.366~M_{\odot}$ with statistical uncertainty $\sigma_{M}=0.003~M_{\odot}$, 
see the red boxes in Fig.~\ref{All_Mg-Mb}.
Meanwhile we know a second object of this kind, the low-mass neutron star companion of 
PSR J1756-2251 \cite{Ferdman:2014rna}
with $M_{G}=1.249 \pm 0.007~M_{\odot}$ which, if it underwent the same evolutionary history as 
PSR~J0737-3039(B) and originated from an O-Ne-Mg star by an electron capture supernova, then it 
should have the same baryon mass, which would be rather troublesome as none of the known 
neutron star EoS candidates would explain such a large gravitational binding energy,  
see the blue boxes in Fig.~\ref{All_Mg-Mb}.   
There are two ways out of this dilemma. Either the mass measurement should get corrected upwards 
or there is another mechanism to create a low-mass neutron star in a binary system which would work for lower baryon masses, such as the so-called "ultra-stripped supernovae" \cite{Tauris:2015xra}. 
Recently, Suwa et al.~\cite{Suwa:2015saa} demonstrated in a simulation the formation of a neutron star with $M_B=1.35~M_\odot$ from such a scenario. 
Conversely, also PSR~J0737-3039(B) could be formed this way. 
Therefore, we derive from the above discussion the following constraint. 
From the two measured neutron star masses discussed above we take the mean value 
$M_{G}=1.24~M_{\odot}$, with the error-bar $\Delta M=\pm0.01~M_{\odot}$, and from both alternative  evolution models we take the mean baryon mass to be $\mu_{B}=1.36~M_{\odot}$ with the error-bar $\Delta M=\pm0.01~M_{\odot}$ (see the grey box in Fig.~\ref{All_Mg-Mb}; the statistical uncertainty is taken to be $\sigma_{M} = \Delta M / 3 = 0.0033~M_{\odot}$.

We compute the conditional probability of the event $E_{P}$ that
the assumed baryon mass is reproducible in the model with parameters $\overrightarrow{\pi}_{i}$
given by normal distribution with mean value $\mu_{B}$. So, using the model
dependance between the masses described by $M_{B}=M_{B}(M_{G};\overrightarrow{\pi}_{i})$
the probability to fulfill the constraint can be calculated from the normal distribution function for that 
$ M_{B}$ values, which are in the area corresponding to the observed gravitational mass range 
$M_{G}\pm\Delta M$:
\begin{eqnarray}
P\left(E_{P}\left|\overrightarrow{\pi}_{i}\right.\right) &=& \Phi(M_{B}(M_{G}+\Delta M;\overrightarrow{\pi}_{i}),\mu_{B},\sigma_{M}) \nonumber\\
&-& \Phi(M_{B}(M_{G}-\Delta M;\overrightarrow{\pi}_{i}),\mu_{B},\sigma_{M}).
\end{eqnarray}
In the plots of results we show the $1\sigma$-range of $ M_{B}$ values as a box on the $M_{G}\otimes M_{B}$ plane.

\subsection{Fictitious radius measurement constraints}

To find out the best suggestion for future observations, which will
be powerful for the model discrimination we employ fictitious
radius measurement constraints. For our "experiment'' we choose
the two known objects with well measured high masses: 
PSR~J$0348+0432$~\cite{Antoniadis:2013pzd}
and 
PSR~J$1614-2230$~\cite{Demorest:2010bx}. 
We assume that the possible radii of these
objects will be different and in the resolution range $\Delta R$.
The masses of the objects are measured $M_{G_{A}}=2.01~M_{\odot}$,
with error-bar $\Delta M_{G_{A}}=\pm0.04~M_{\odot}$ and $M_{G_{D}}=1.94~M_{\odot}$,
with error-bar $\Delta M_{G_{D}}=\pm0.04~M_{\odot}$ correspondingly
for the PSR~J0348+0432 and PSR~J1614-2230 pulsars. 
The fictitious radii we assumed to be $R_{A}$ and $R_{D}$ with
statistical uncertainty $\sigma_{R_{A}}$ and $\sigma_{R_{D}}$ correspondingly.

We compute the conditional probability of the event $E_{F}$  for fixed model with parameters $\overrightarrow{\pi}_{i}$ under the assumption that
the "measured" radii should be producible with normal distribution with mean values 
$R_{A}$ and $R_{D}$. Each
model defines dependence of the radii on masses for all
possible stable branches (families of neutron stars) given by $R_{\alpha}=R_{\alpha}(M_{G_{\alpha}};\overrightarrow{\pi}_{i})$
functions. 
Here the index $\alpha$ stands for the definition of the pulsar. 
The probability to fulfil the constraint will correspond to the area where the configurations
have values of radii predicted from the model when the masses are in the interval of the observational mass range $M_{G_{j}}\pm\Delta M_{j}$: 
\begin{eqnarray}
P\left(E_{F_{\alpha}}\left|\overrightarrow{\pi}_{i}\right.\right) &=& \Phi\left(R_{\alpha}(M_{G_{\alpha}}+\Delta M_{\alpha};\overrightarrow{\pi}_{i}),\mu_{R_{\alpha}},\sigma_{R_{\alpha}}\right) \nonumber\\
&-& \Phi\left(R_{\alpha}(M_{G_{j}}-\Delta M_{\alpha};\overrightarrow{\pi}_{i}),\mu_{R_{\alpha}},\sigma_{R_{\alpha}}\right)
\label{p_bogdan-2-1}
\end{eqnarray}
Because in some cases (mainly for the hybrid stars) the $R_{\alpha}(M_{G_{\alpha}};\overrightarrow{\pi}_{i})$
can be not uniquely defined functions, we have excluded possible overlaps of
the boxes of the radius probability
regions on the $M_{G_{\alpha}}\otimes R_{\alpha}$ plane to avoid the double counting.

\subsection{Calculation of \textit{a posteriori} probabilities}

It is important to note that these measurements are independent of
each other. This means that we can compute the complete conditional
probability of an event $E$ given $\overrightarrow{\pi}_{i}$ that
corresponds to the product of the conditional probabilities of all
measurements, in our case resulting from the constraints on maximum mass $E_{A}$,
on maximum radii $E_{B}$, from gravitational mass - baryon mass relation  $E_{P}$ and for 
fictitious radii measurements for two objects $E_{F_A}$ and $E_{F_D}$.   
\begin{equation}
P\left(E\left|\overrightarrow{\pi}_{i}\right.\right)= \prod_{\alpha} P\left(E_{\alpha}\left|\overrightarrow{\pi}_{i}\right.\right).
\label{p_event}
\end{equation}
Thus, we can derive the probability of the measurement of an EoS represented
by a vector of parameters $\overrightarrow{\pi}_{i}$ using Bayes' theorem
\begin{equation}
P\left(\overrightarrow{\pi}_{i}\left|E\right.\right)=\frac{P\left(E\left|\overrightarrow{\pi}_{i}\right.\right)P\left(\overrightarrow{\pi}_{i}\right)}{\sum\limits _{j=0}^{N-1}P\left(E\left|\overrightarrow{\pi}_{j}\right.\right)P\left(\overrightarrow{\pi}_{j}\right)}.\label{pi_apost}
\end{equation}
In this article we consider BA for three different groups of constrains (cases) as follows  
\begin{eqnarray*}
       {\rm Case~ 1:~}  \alpha&=&{\rm A, B;} \\
       {\rm Case~ 2:~}  \alpha&=&{\rm A, B, P; }\\
       {\rm Case~ 3:~ } \alpha&=&{\rm F_A, F_D. }
\end{eqnarray*}

\begin{figure*}[!htpb]
\begin{center}
\begin{tabular}{l|c|c|c}
\hline
Radius error $\sigma \to$ &&&\\[-2mm]
&1.5 km& 1.0 km& 0.5 km\\[-2mm]
$R_A , R_D$&&&\\
\hline
&&&\\[-2mm] 
11 km, 13 km&
\includegraphics[width=0.25\textwidth]{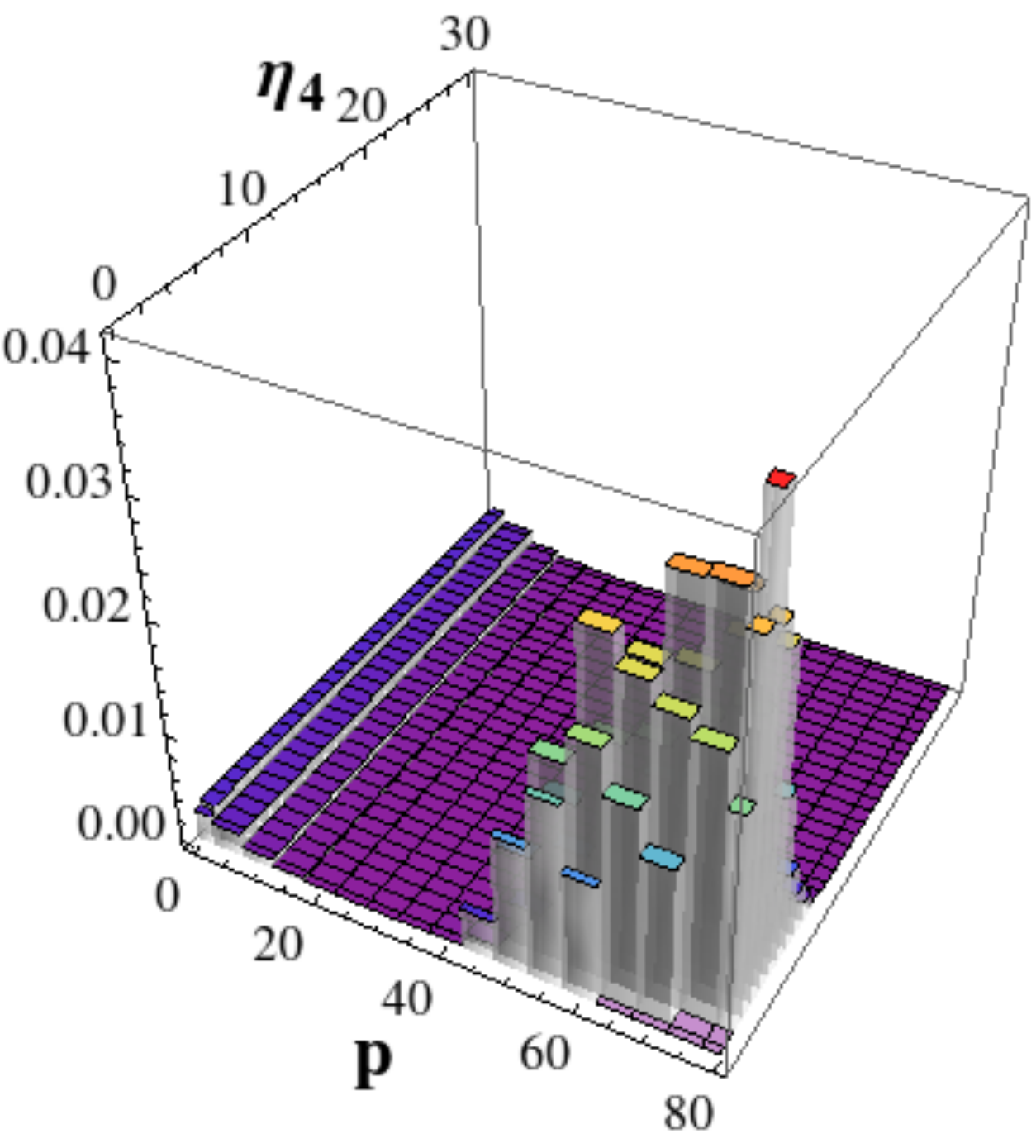} & 
\includegraphics[width=0.25\textwidth]{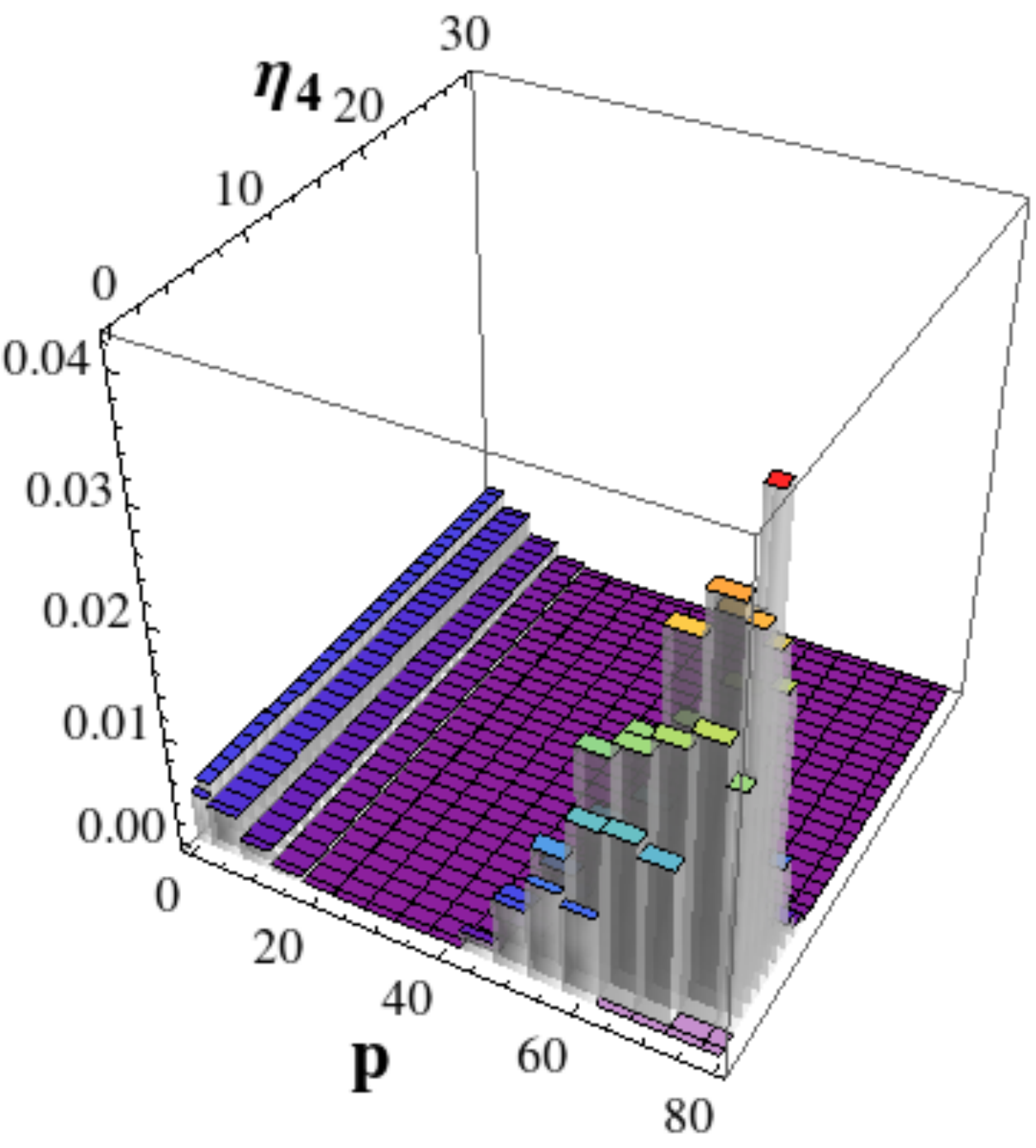} & 
\includegraphics[width=0.25\textwidth]{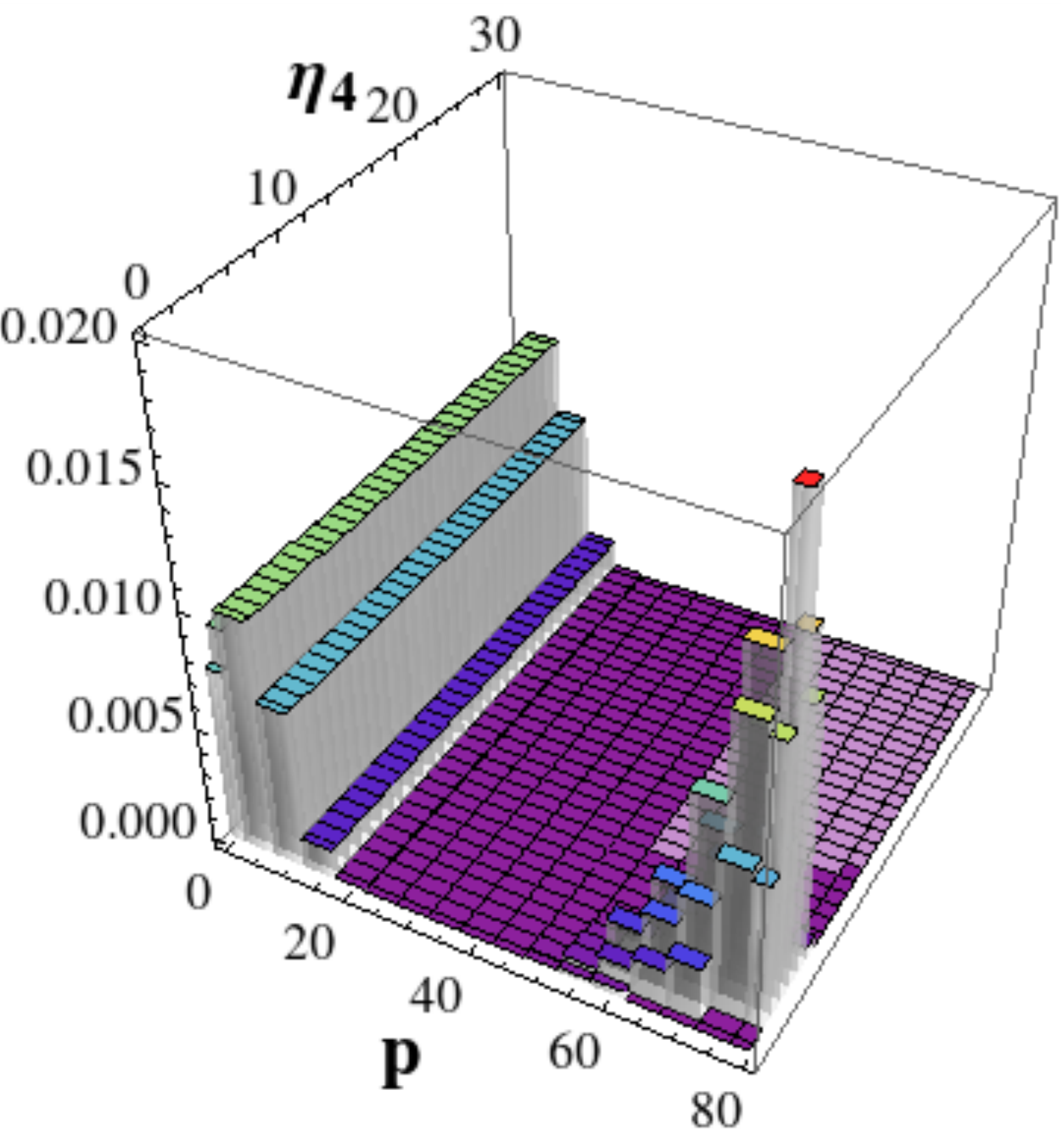} \\
&&&\\[-2mm]
\hline
&&&\\ [-2mm]
11 km, 15 km&
\includegraphics[width=0.25\textwidth]{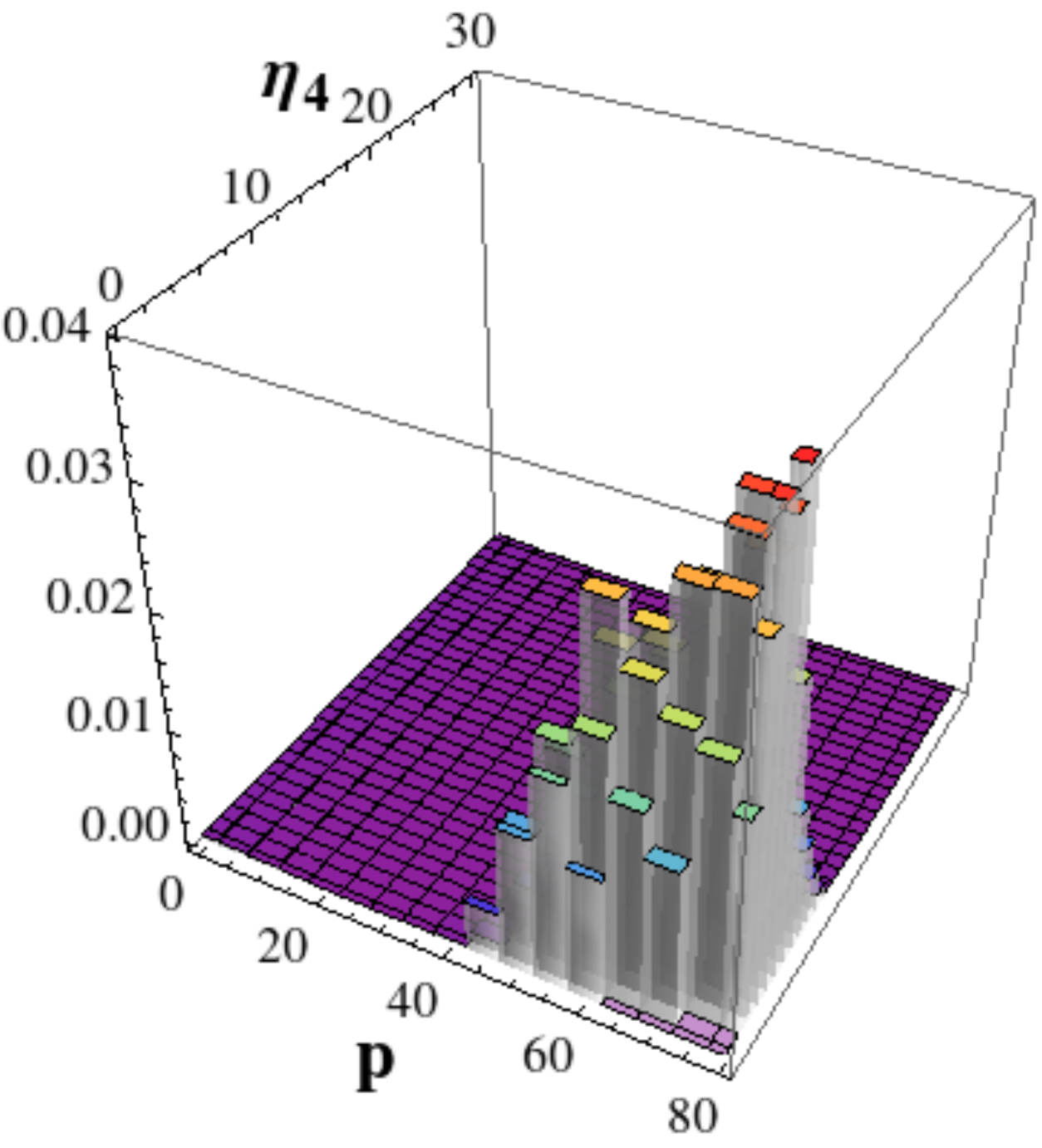} & 
\includegraphics[width=0.25\textwidth]{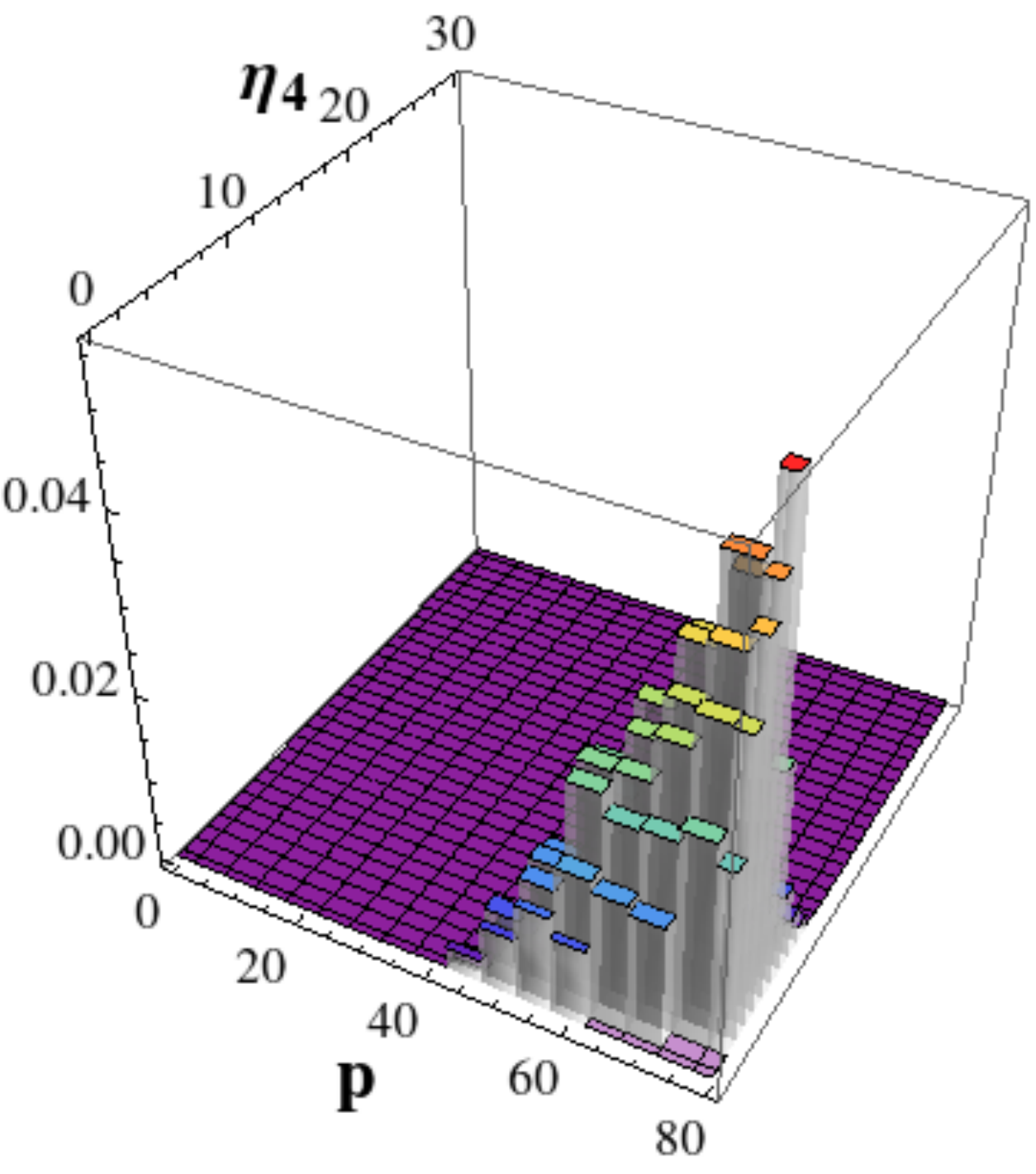} & 
\includegraphics[width=0.25\textwidth]{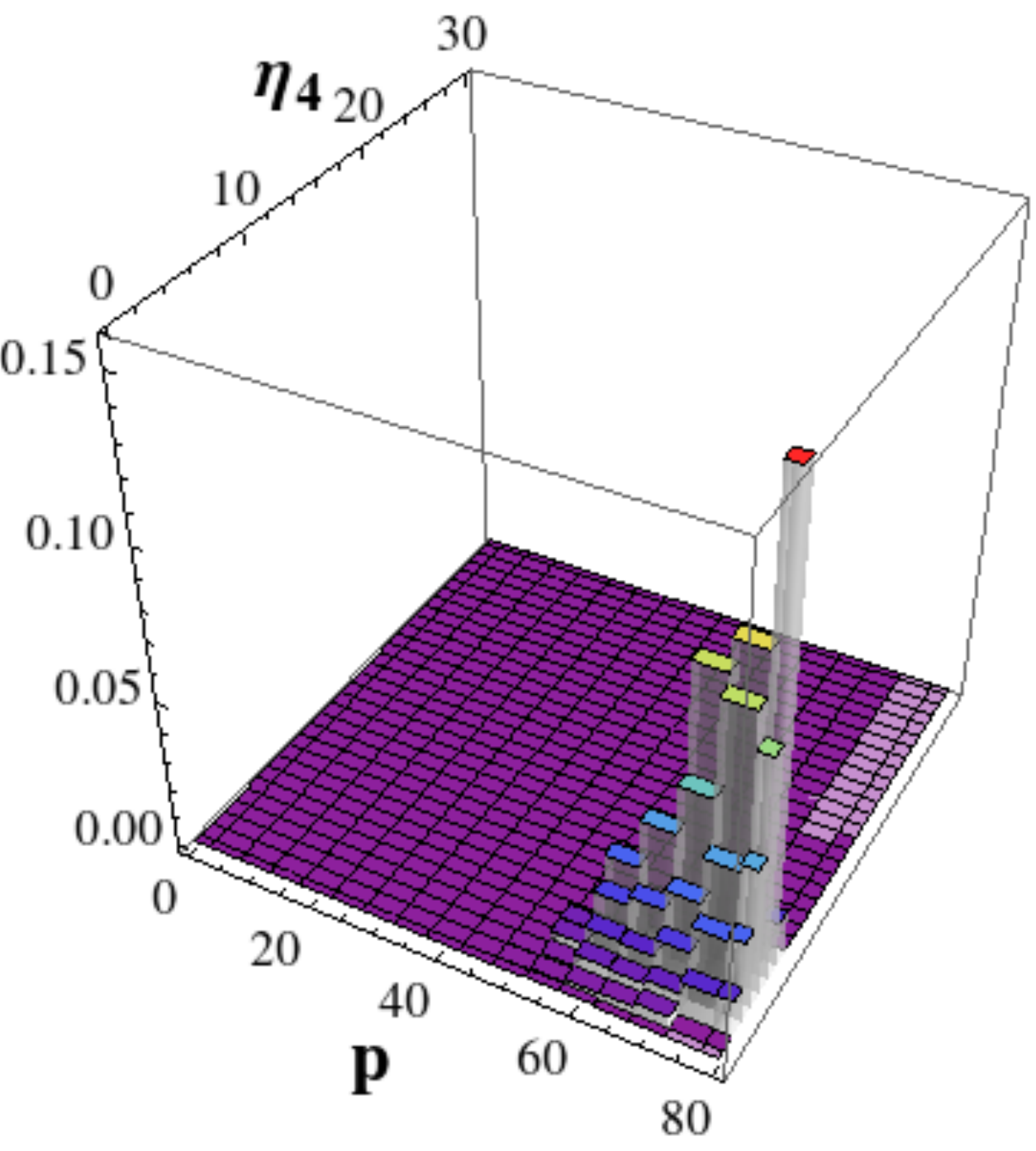} \\
&&&\\[-2mm]
\hline
&&&\\[-2mm] 
13 km, 13 km&
\includegraphics[width=0.25\textwidth]{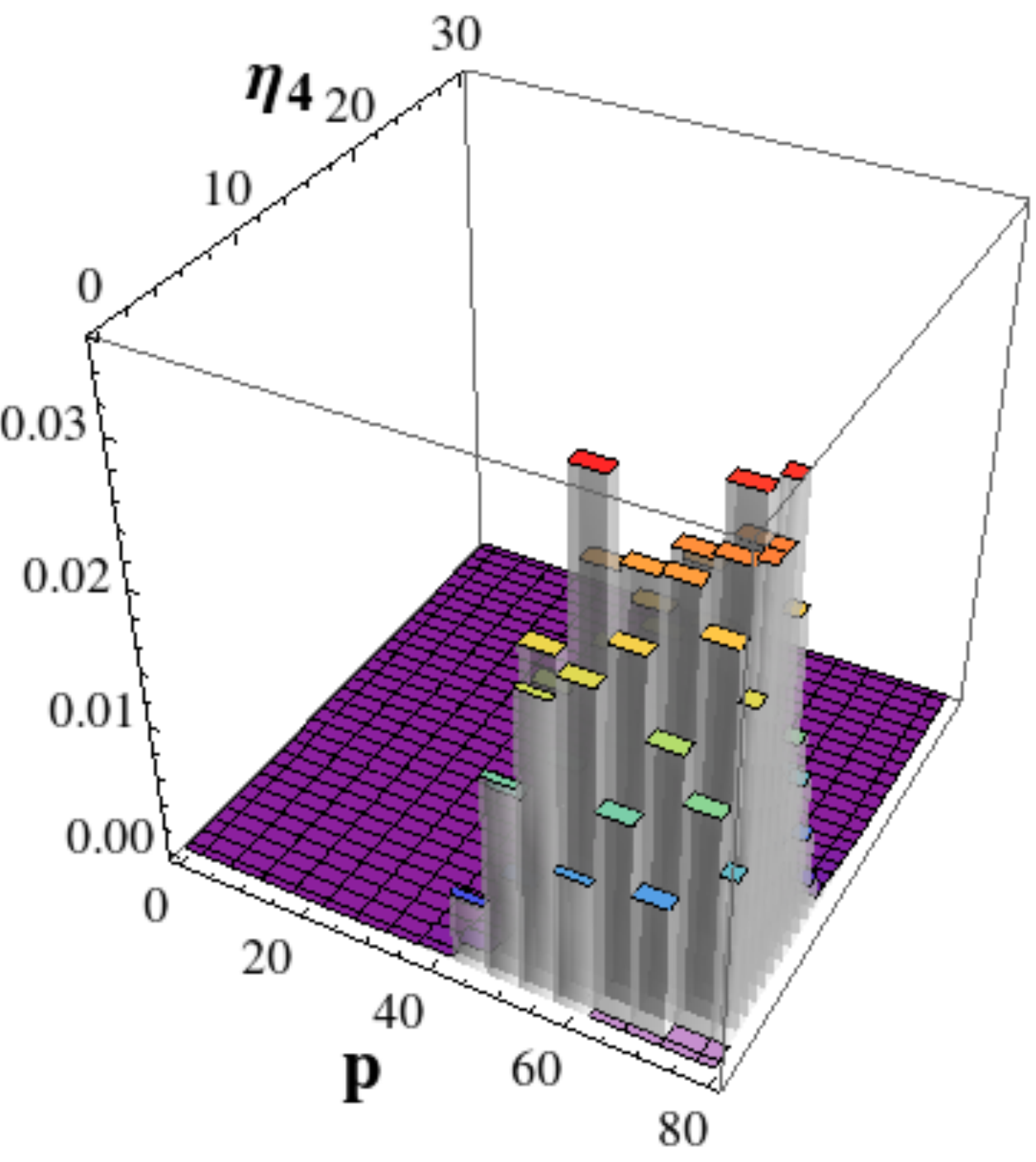} & 
\includegraphics[width=0.25\textwidth]{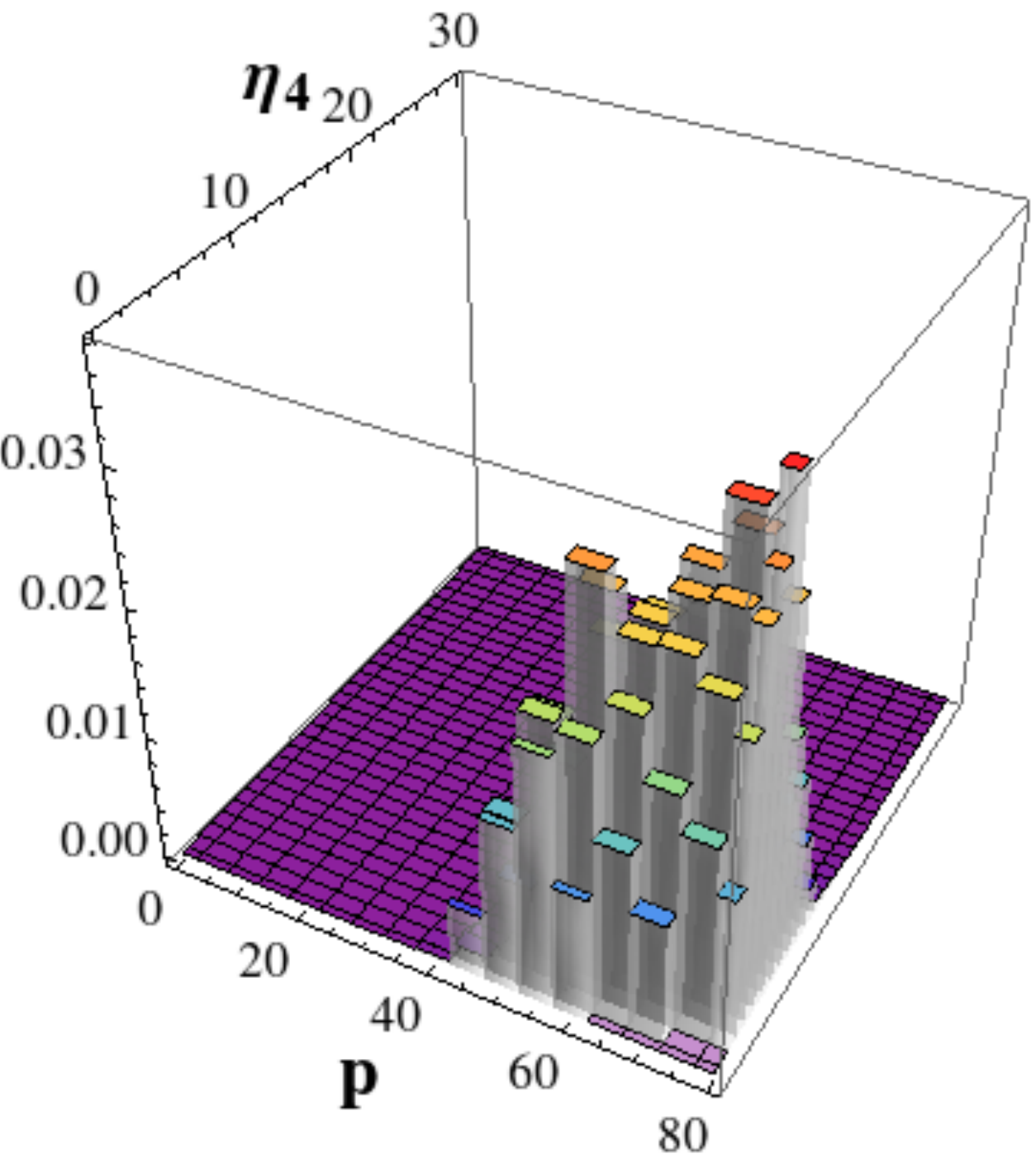} & 
\includegraphics[width=0.25\textwidth]{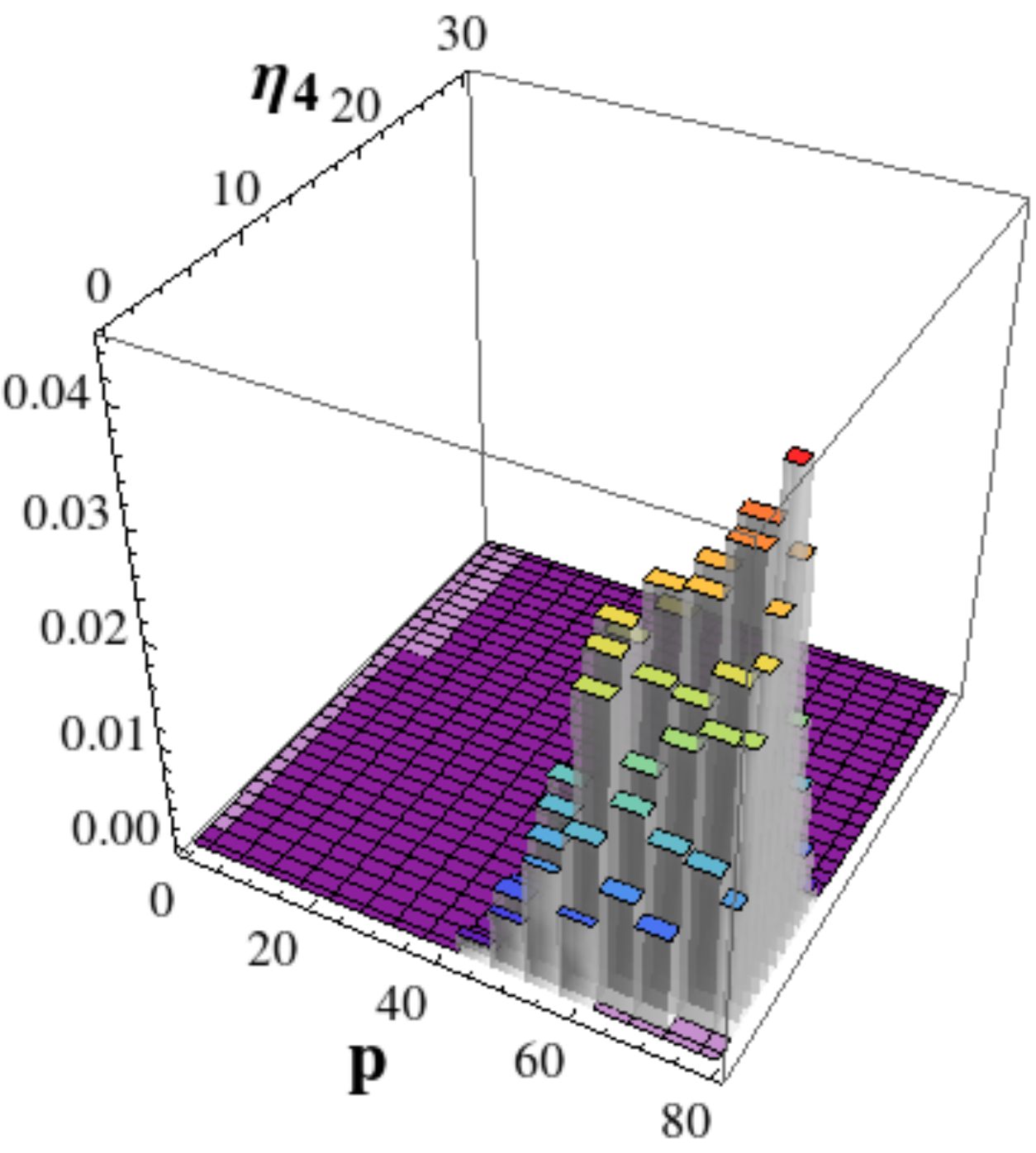}\\
&&&\\[-2mm]
\hline
&&&\\[-2mm] 
15 km, 11 km&
\includegraphics[width=0.25\textwidth]{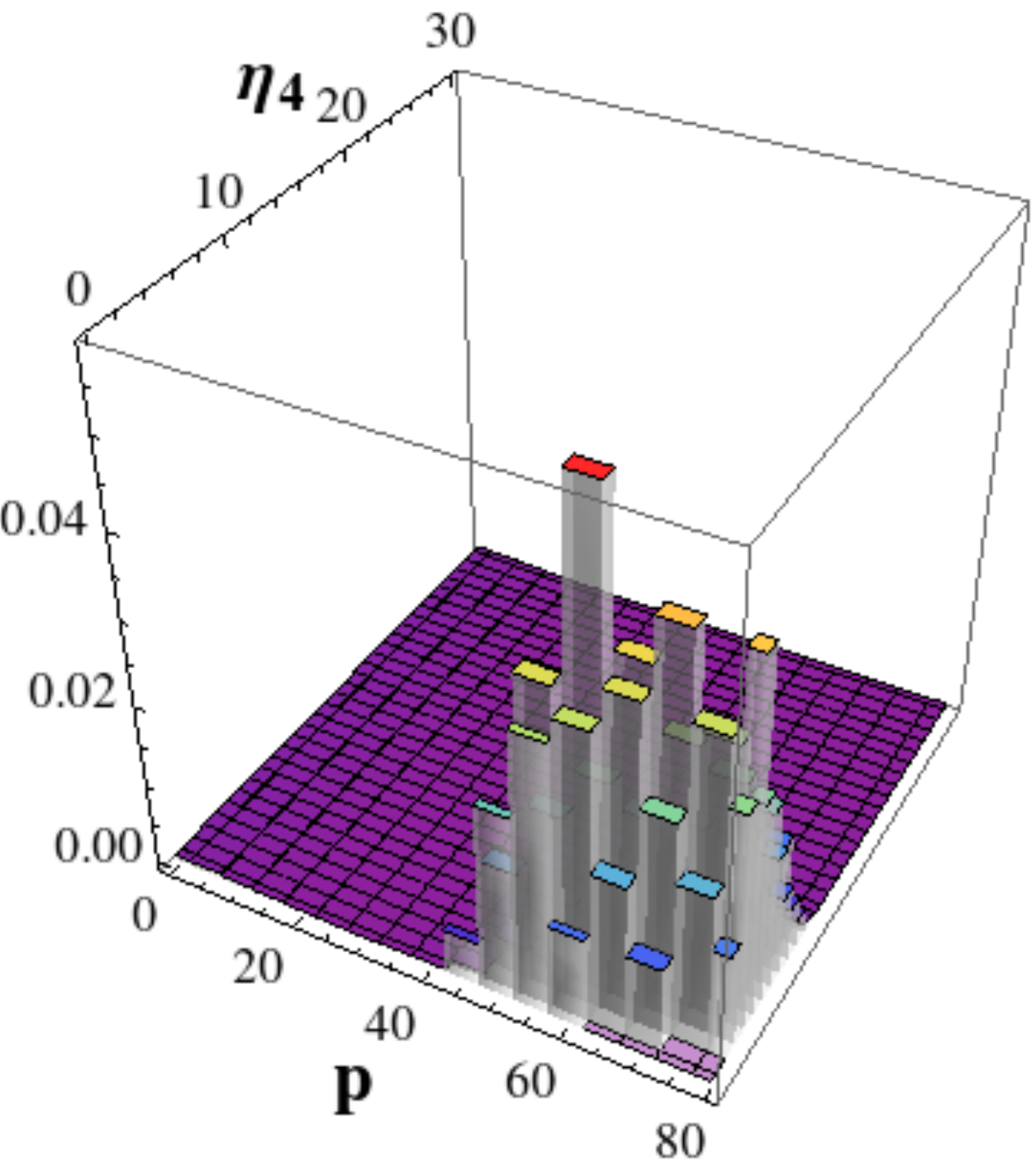} & 
\includegraphics[width=0.25\textwidth]{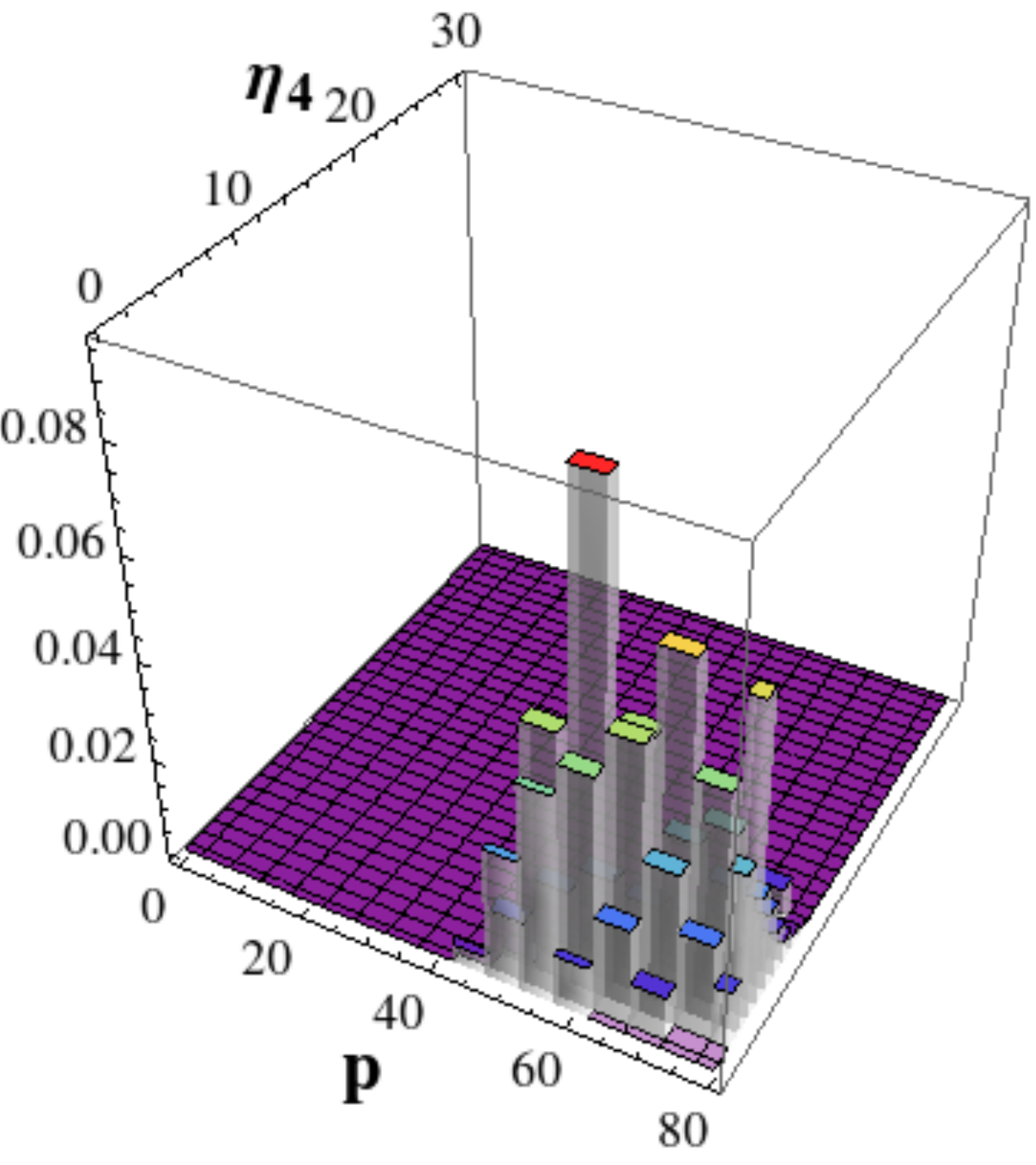} & 
\includegraphics[width=0.25\textwidth]{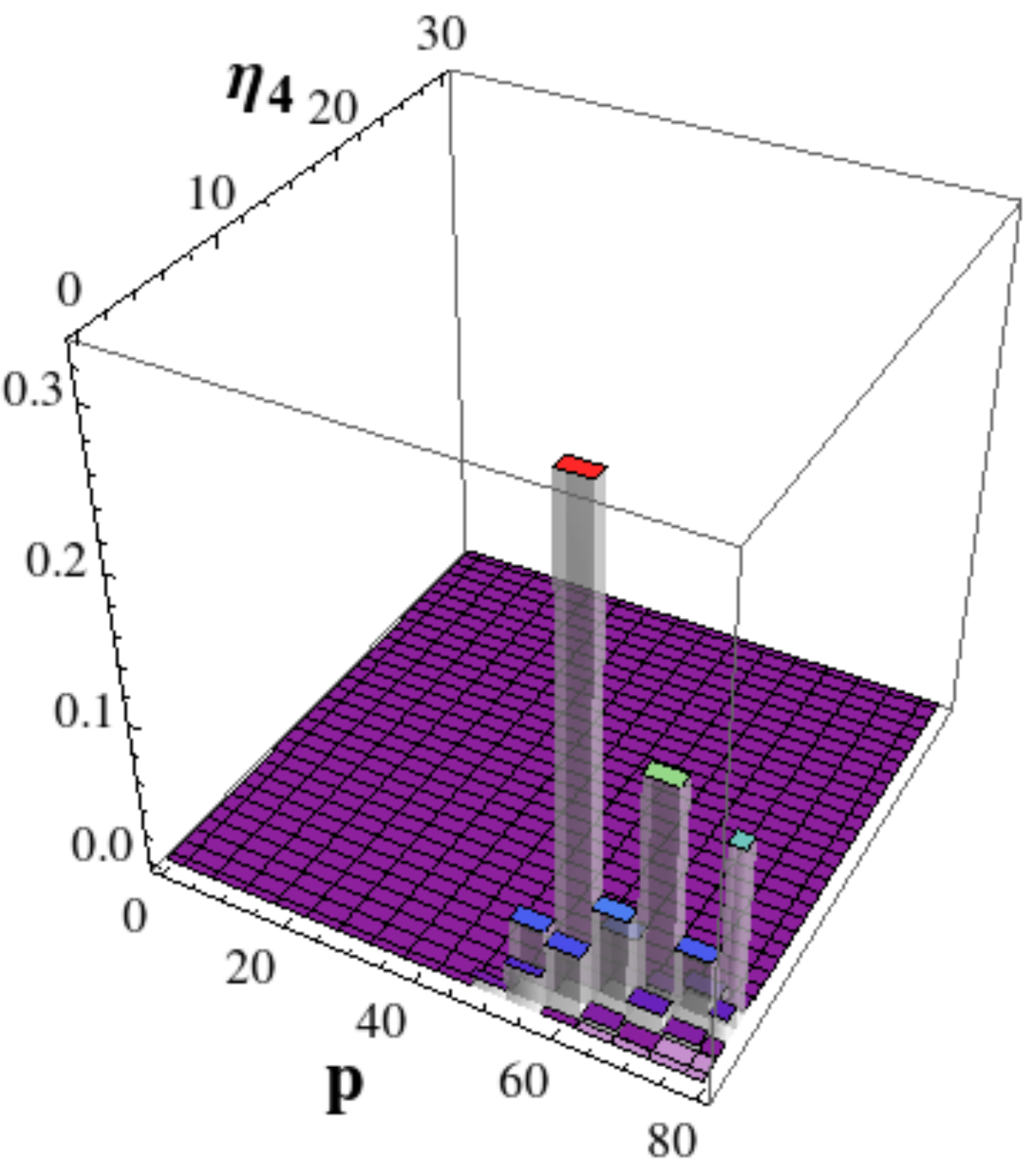}\\
&&&\\
\hline 
\end{tabular}
\end{center}
\caption{\textit{Probabilities for an extra fictitious radius measurement}. 
$R_A$ and $R_D$ denote NS with masses corresponding to the ones measured by Antoniadis et al. and by Demorest et al., resp.}
\label{Prob_Fic}
\end{figure*}

For the Case~3 we have chosen different sets of fictitious radius measurements (see Fig.~\ref{Prob_Fic}), 
assuming that the more massive object PSR~J0348+0432~\cite{Antoniadis:2013pzd} should have the smaller radius and, therefore, PSR~J1614-2230~\cite{Demorest:2010bx} has the larger fictitious radius. 
Anyway, we also consider one case (last row of Fig.~\ref{Prob_Fic}) where the less massive pulsar has the lower radius.

\section{Results}
\label{sec:results}

In order to cover a relevant set of possible hybrid compact star EoS in the pressure-energy density plane
we have varied the excluded parameter in the range $p= 0,5, \dots, 80$ 
as well as the 8-quark vector current coupling in the range $\eta_4=0,1,2,\dots, 30$.

The hybrid EoS resulting from Maxwell constructions between all combinations of hadronic and quark
matter EoS in this two-dimensional parameter space is shown in Fig.~\ref{All_M-R} together with the corresponding NS sequences.

Note that compact star sequences in the mass-radius plane which have a vertical branch correspond to hadronic stars and an almost horizontal branch corresponds to 
hybrid stars with a quark matter core. 
The parameters are chosen in such a range that their variation entails that the hybrid star branch 
becomes disconnected from the hadronic one due to the appearance of a set of unstable configurations.
This characterizes the appearance of high-mass twin stars and happens in particular when varying the the excluded volume value in all of these models.
We have already observed in \cite{Benic:2014jia} that increasing the $\eta_4$ parameter for a fixed $p$ value corresponds to an increase of the maximum mass, 
while the difference between the radius of the hybrid star and its hadronic twin decreases. 
This general behaviour is preserved for any fixed $p$.

The general behavior of the symmetry energy in each class of EoS is to shift radius values while preserving the maximum mass for each sequence. Even though  the general result
is that the DD2F class presents lower average radii NS than the DD2, the symmetry energy and the excluded volume can modify the average radius values. 
The DD2+ class is therefore
the stiffest EoS leading to larger radii whereas the DD2F- shows the lowest radii by being the softest EoS class. 

{\it A posteriori} probabilities of identification for these twin EoS are shown in figures ~\ref{First_Lego} and \ref{Second_Lego}. 
The models with higher probabilities correspond to the ones that predict
large radii and are associated to large excluded volume parameters $p$.
The optimal value of the vector coupling strength is $\eta_4=5$.
We note that the family of hybrid EoS accessible within the given two-dimensional parameter space 
describes sequences of compact stars that have a hadronic branch and a hybrid star branch.
The latter appears connected to the former for not too large excluded volumina.
For sufficiently large excluded volume parameters the compact stars on the hadronic branch have radii 
exceeding $14$ km and the phase transition proceeds with a sufficiently large jump in the energy density 
so that the hybrid star sequence gets disconnected from the hadronic one, forming a so-called "third family". 
The observation of a corresponding (almost) horizontal branch in the $M-R$ diagram would be a clear signal 
for a strong first order phase transition in the high-density neutron star EoS. 
Using Bayesian analysis technique we have demonstrated that the measurement of significantly different radii 
for two high-mass compact stars (like the $2~M_\odot$ objects we already know)  would have sufficient discriminating power to favor a hybrid EoS over a purely hadronic one.    


\section{Conclusions}
\label{sec:conclusion}

We summarize our conclusions from this study of a set of six classes of two-parameter hybrid EoS for compact star interiors:
\begin{enumerate}
\item The most probable models exhibit high-mass twin star configurations with quite distinguishable radii,
differing by about 2 km.
\item The region of the most probable models in the two-dimensional parameter space is sufficiently narrow, covering the ranges $40<p < 80$ and $3<\eta_4< 7$.
\item The most probable models have a large excluded volume parameter $p> 40$ and a 
not too large vector coupling strength $\eta_4\sim 5$.
\item The existence of the horizontal branch signals a strong first order deconfinement phase transition and is a feature accessible to verification by observation. 
To that end, at least for two high-mass pulsars with masses $\sim 2~M_\odot$ (like PSR J1614-2230 and PSR J0348+0432) the radii should be measured to sufficient accuracy and turn out to be significantly different.
\item Fig.~\ref{Prob_Fic} shows that there are strong peaks of probabilities in the parameter space, even if fictitious radius measurements have quite large uncertainties with 
$\sigma_{\rm R_A} = \sigma_{\rm R_D} = 1.5$ km, where the ''measurements'' have overlap of $5\sigma$ regions. It means that even quite uncertain measurements of radii for massive pulsars could have a strong selective power of EoS models. Cf. cases 1) and 2) which demonstrate that knowing only either mass or radius of an object results in a constraint which is not very robust against a possible third observational constraint, like the gravitational binding energy. 
\item When fictitious radius measurements yield the smaller radius for the object with the slightly smaller mass
(see the bottom line of Fig.~\ref{Prob_Fic}), then the most probable value of the excluded volume parameter is lowered from $p= 80$ to the moderate $p\sim 50$, while the optimal stiffness of the quark matter remains unchanged, $\eta_4\sim 5$.  
\end{enumerate}
The next two steps in the development of the approach are devoted to an improvement of the variability of the dense matter EoS within a two-dimensional parameter space embodying, e.g., also the purely hadronic case without a phase transition and to mimicking the occurrence of structures (so-called "pasta phases") in the phase transition region \cite{Yasutake:2014oxa,Alvarez-Castillo:2014dva}.

%
%
%
%

\subsection*{Acknowledgements}
We acknowledge the partial support by the Helmholtz Association (HGF) through the Nuclear Astrophysics Virtual Institute (VH-VI-417) and by the COST Action MP 1304 "NewCompStar" for our international 
networking activities in preparing this article.
This work received support from the Polish NCN under grant No. UMO-2014/13/ B/ST9/02621.
D.E.A-C. and H.G. are grateful for support from the programme for exchange between JINR Dubna and 
Polish Institutes (Bogoliubov-Infeld programme). 
\\
D.E.A-\-C. and S.T. received support  form the Heisenberg-Landau programme for scientist exchange between JINR Dubna and German Institutes.
S. B. acknowledges partial support by the Croatian Science Foundation under Project No. 8799. 
D.B. was supported in part by the Hessian LOEWE initiative through HIC for FAIR.


\end{document}